\documentclass[a4paper,11pt]{article}
\pdfoutput=1

\usepackage{jheppub}
\usepackage{graphicx}
\usepackage{amsmath}

\newcommand{\eq}[1]{eq.~\eqref{eq:#1}}

\renewcommand{\sec}[1]{section~\ref{sec:#1}}

\newcommand{\app}[1]{Appendix~\ref{app:#1}}
\newcommand{\fig}[1]{figure~\ref{fig:#1}}

\newcommand{\mycites}[1]{refs.~\cite{#1}}
\newcommand{\mycite}[1]{ref.~\cite{#1}}

\newcommand{\lqcd}{\Lambda_\mathrm{QCD}}

\allowdisplaybreaks[2]

\newcommand{\abs}[1]{\lvert#1\rvert}

\newcommand{\ord}[1]{{\mathcal O}(#1)}

\newcommand{\ORD}[1]{{\mathcal O}\biggl(#1\biggr)}

\newcommand{\Mae}[3]{\bigl\langle#1\bigr\rvert#2\bigr\rvert#3\bigr\rangle}

\newcommand{\bare}{\mathrm{bare}}

\newcommand{\bn}{\bar{n}}

\newcommand{\df}{\mathrm{d}}
\newcommand{\img}{\mathrm{i}}

\newcommand{\Li}{\mathrm{Li}}

\newcommand{\eps}{\epsilon}
\newcommand{\w}{\omega}

\newcommand{\cB}{{\mathcal B}}
\newcommand{\cL}{{\mathcal L}}
\newcommand{\cI}{{\mathcal I}}

\newcommand{\tB}{\widetilde{B}}

\newcommand{\GeV}{\,\mathrm{GeV}}

\newcommand{\nn}{\nonumber}

\newcommand{\hp}{\hat{p}}
\newcommand{\bnP}{\overline {\mathcal P}}

\newcommand{\conv}{\!\otimes\!}
\newcommand{\convz}{\!\otimes_z\!}

\newcommand{\zero}{{(0)}}
\newcommand{\one}{{(1)}}
\newcommand{\two}{{(2)}}

\newcommand{\tr}{\mathrm{tr}}

\newcommand{\cusp}{\mathrm{cusp}}

\title{The Gluon Beam Function at Two Loops}

\author{Jonathan R.~Gaunt,}
\author{Maximilian Stahlhofen}
\author{and Frank J.~Tackmann}

\affiliation{Theory Group, Deutsches Elektronen-Synchrotron (DESY), Notkestra\ss e 85, D-22607 Hamburg, Germany}

\emailAdd{jonathan.gaunt@desy.de}
\emailAdd{maximilian.stahlhofen@desy.de}
\emailAdd{frank.tackmann@desy.de}

\abstract{
The virtuality-dependent beam function is a universal ingredient in the resummation for observables probing the virtuality of incoming partons, including $N$-jettiness and beam thrust. We compute the gluon beam function at two-loop order. Together with our previous results for the two-loop quark beam function, this completes the full set of virtuality-dependent beam functions at next-to-next-to-leading order (NNLO). Our results are required to account for all collinear initial-state radiation effects on the $N$-jettiness event shape through N$^3$LL order. We present numerical results for both the quark and gluon beam functions up to NNLO and N$^3$LL order. Numerically, the NNLO matching corrections are important. They reduce the residual matching scale dependence in the resummed beam function by about a factor of two.
}


\begin{document}
{\flushright DESY 14-004\\May 05, 2014\\[-9ex]}
\maketitle

\section{Introduction}
\label{sec:intro}

In differential measurements at hadron colliders, collinear initial-state radiation is described and can be resummed by process-independent beam functions~\cite{Stewart:2009yx}. In this paper, we are concerned with the two-loop virtuality-dependent beam functions $B_i(t,x,\mu)$, which are defined formally as the proton matrix element of operators in soft collinear effective field theory (SCET)~\cite{Bauer:2000ew, Bauer:2000yr, Bauer:2001ct, Bauer:2001yt, Bauer:2002nz, Beneke:2002ph} in \mycites{Stewart:2009yx, Stewart:2010qs}.

The beam functions $B_i(t,x,\mu)$ are an integral component for predictions of observables in hadronic processes that probe the virtuality of the incoming partons via a measurement performed on the hadronic final state, such as $N$-jettiness \cite{Stewart:2010tn} and beam thrust \cite{Stewart:2009yx}.
The $B_i(t,x,\mu)$ are a type of unintegrated parton density that, loosely speaking,
gives the probability to find a parton $i$ in the initial state carrying a fraction $x$ of the proton's lightcone
momentum, accompanied by initial-state radiation that causes the parton $i$ to be off shell with virtuality $-t$.%
\footnote{Another example of a beam function that recently has reached two-loop precision~\cite{Gehrmann:2012ze, Catani:2013tia, Gehrmann:2014yya} is the transverse momentum dependent PDF (TMD PDF), which is relevant when the total transverse momentum of the hard final state is measured.}

The SCET definition of the virtuality-dependent gluon beam function ($i=g$) relevant for this work reads ($x\equiv \w/P^-$)
\begin{align} \label{eq:Bg_def}
B_g(t, x, \mu) &= \Mae{p_n(P^-)}{(-\w)\theta(\w) \, \cB_{n\perp\mu}^c(0)\, \delta(t - \w\hp^+) \bigl[\delta(\w - \bnP_n) \cB_{n\perp}^{\mu c}(0) \bigr]}{p_n(P^-)}
\,,\end{align}
where $p_n(P^-)$ denotes the incoming (spin-averaged) proton state with lightlike momentum $P^\mu=P^- n^\mu/2$, $\bnP_n$ is the SCET minus-momentum label operator~\cite{Bauer:2001ct}, $\hp^+$ is the plus-momentum operator, and $\cB_{n\perp}^\mu$ denotes the gauge-invariant gluon field strength operator in SCET. For more details on the SCET notations and conventions, we refer to \mycites{Stewart:2009yx, Stewart:2010qs, Gaunt:2014xga}.

For $t \gg \Lambda_\mathrm{QCD}$, the beam function can be obtained as the convolution of the collinear parton
density functions (PDFs) $f_i(x,\mu)$ with perturbatively calculable matching functions $\cI_{ij}(t,z,\mu)$~\cite{Fleming:2006cd, Stewart:2009yx, Stewart:2010qs}
\begin{align} \label{eq:BOPE}
B_{i}(t,x,\mu) &= \sum_j  \int^1_x \! \dfrac{\df z}{z}\, \cI_{ij}(t,z,\mu)\, f_{j}\Bigl(\frac{x}{z},\mu \Bigr)
\bigg[1 + \ORD{\frac{\lqcd^2}{t}} \bigg]
\,.\end{align}
The matching coefficients were calculated to one-loop order for the quark case ($i=q$) in \mycite{Stewart:2010qs}
and for the gluon case ($i=g$) in \mycite{Berger:2010xi}. In a previous publication \cite{Gaunt:2014xga} we calculated
the (anti)quark matching coefficients at two-loop order (NNLO). In this paper, we complete the full set
of matching coefficients through NNLO.

In \sec{match}, we give a brief summary of the calculational approach, referring to \mycite{Gaunt:2014xga} for more details on the setup of the matching calculation. In the appendices, we give several technical details on the calculation.
In \sec{results}, we present our results for the gluon matching coefficients at two loops.
In \sec{plots}, we present numerical results for both the quark and gluon beam functions up to NNLO and N$^3$LL order.
We conclude in \sec{conclusions}.

\section{Calculation}
\label{sec:match}

\begin{figure}[t]
\includegraphics[width=0.245\textwidth]{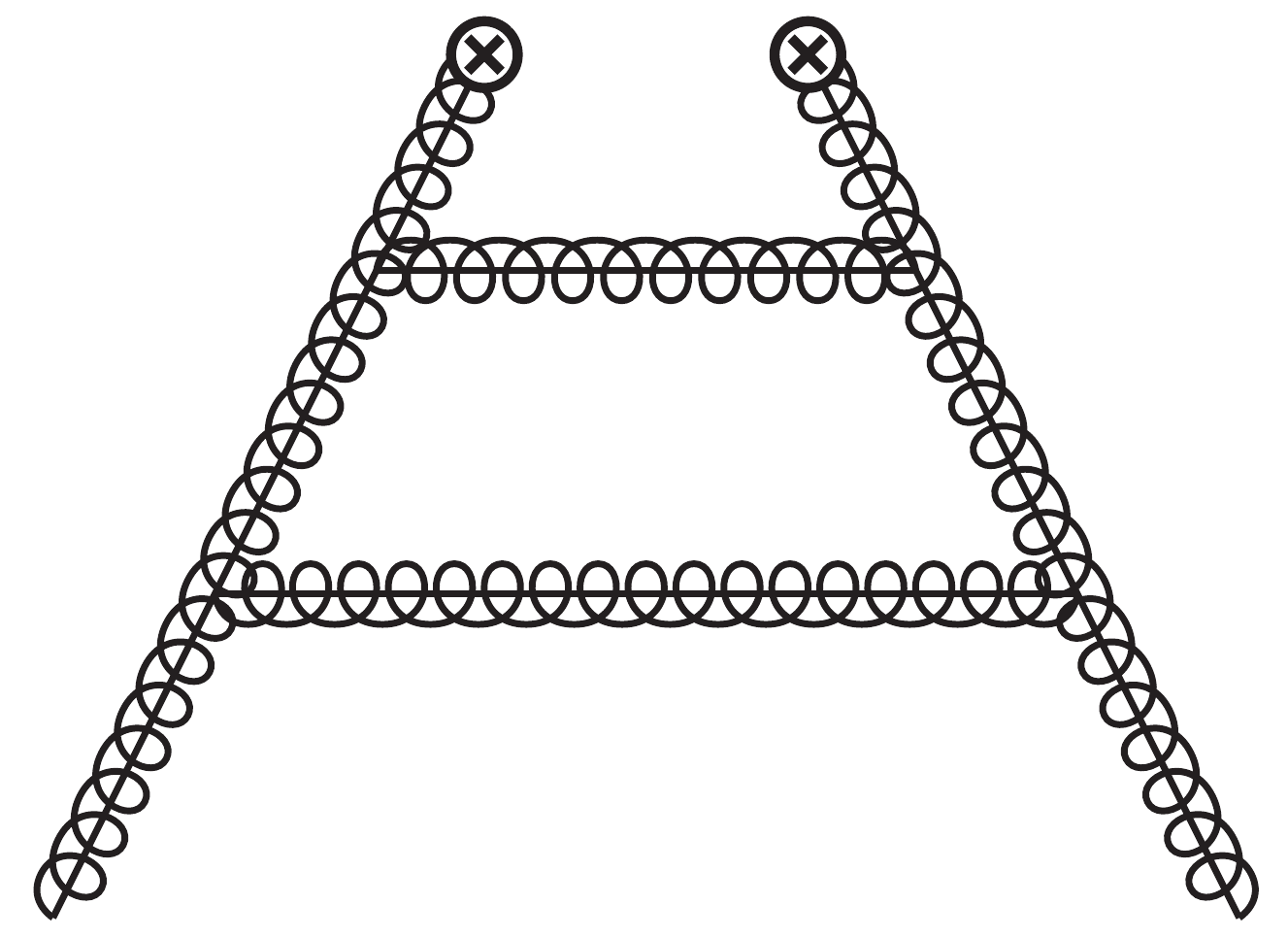}
\put(-100,68){a)}
\includegraphics[width=0.245\textwidth]{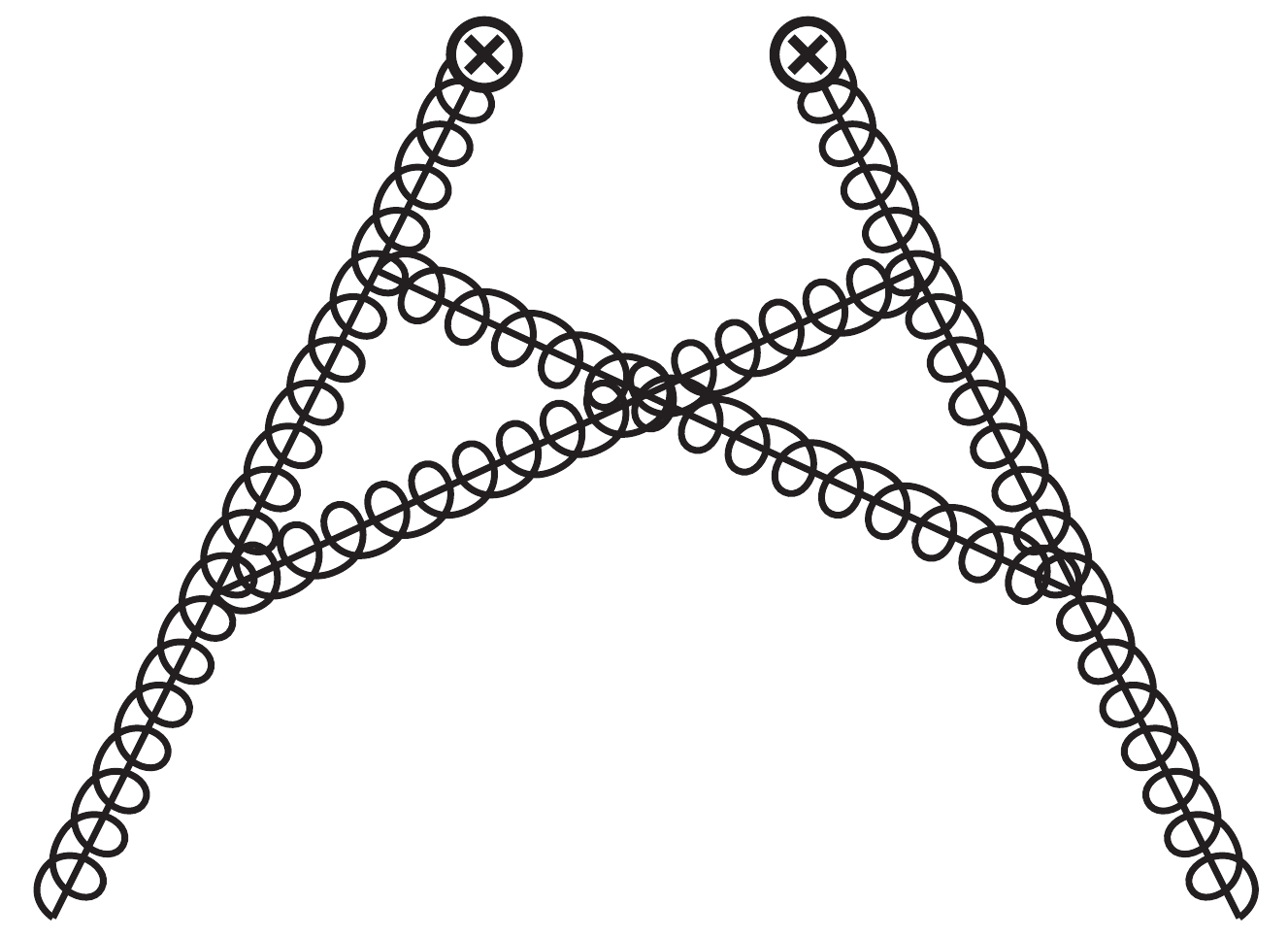}
\put(-100,68){b)}
\includegraphics[width=0.245\textwidth]{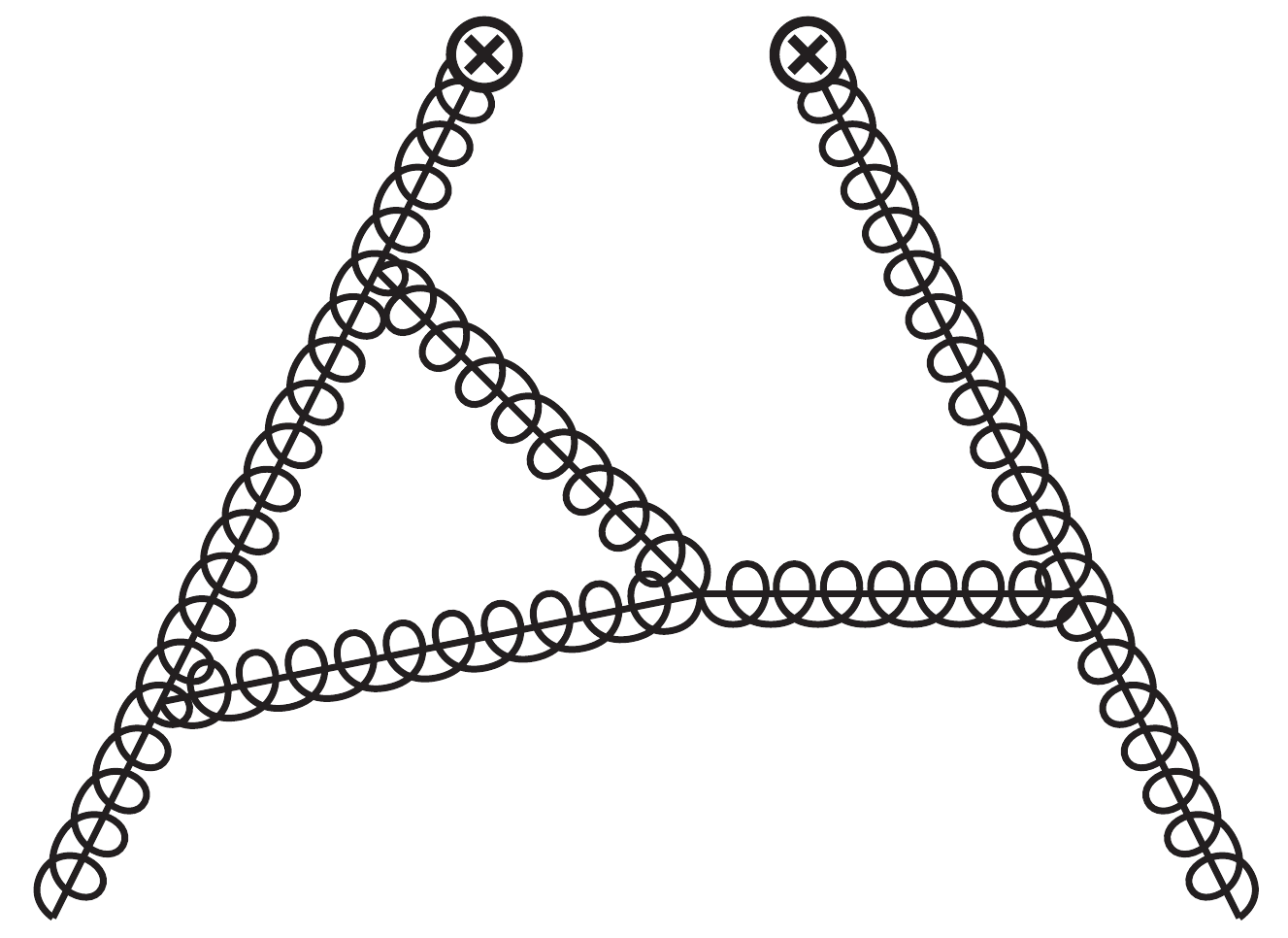}
\put(-100,68){c)}
\includegraphics[width=0.245\textwidth]{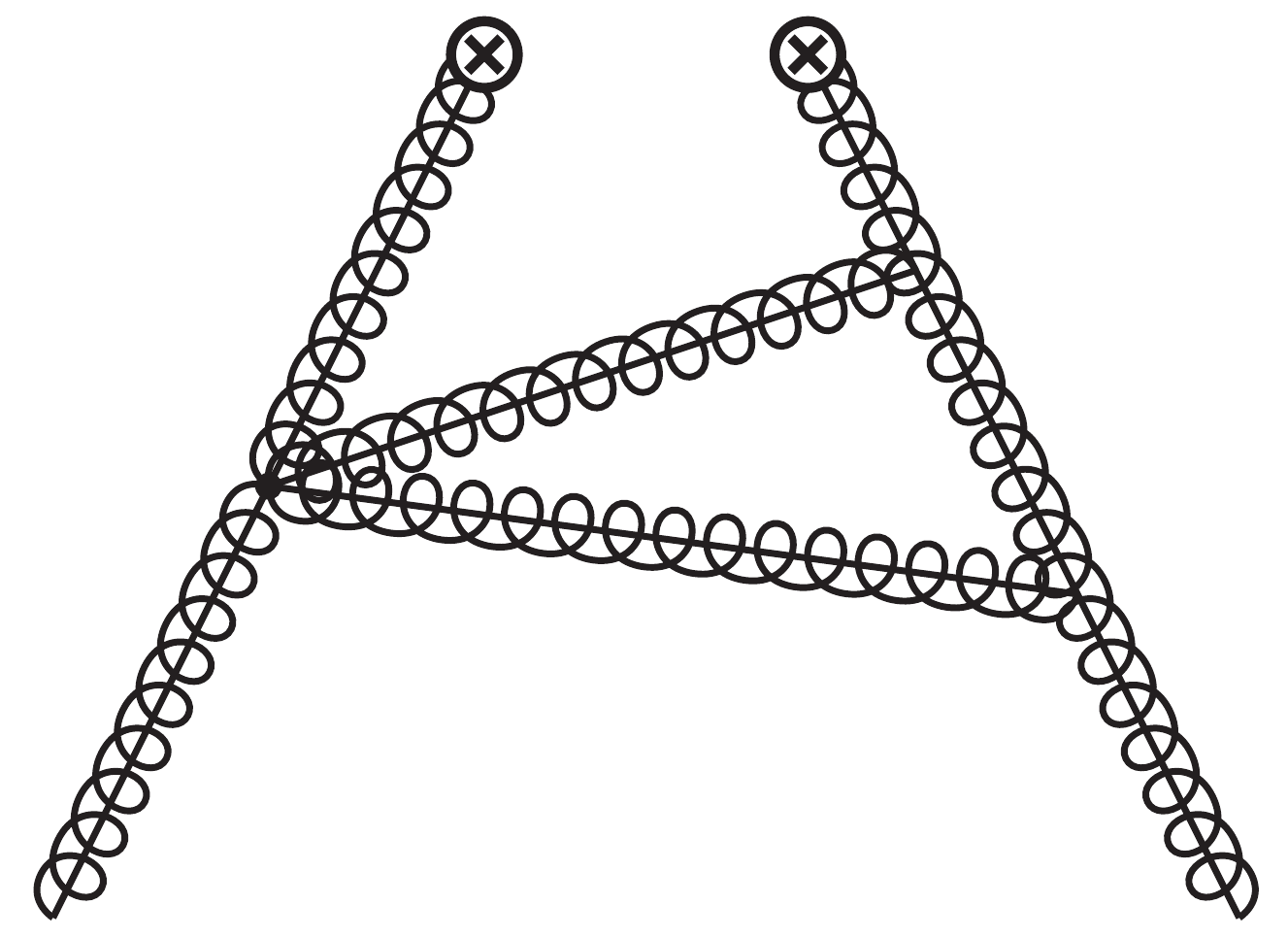}
\put(-100,68){d)}
\vspace{1ex}
\includegraphics[width=0.245\textwidth]{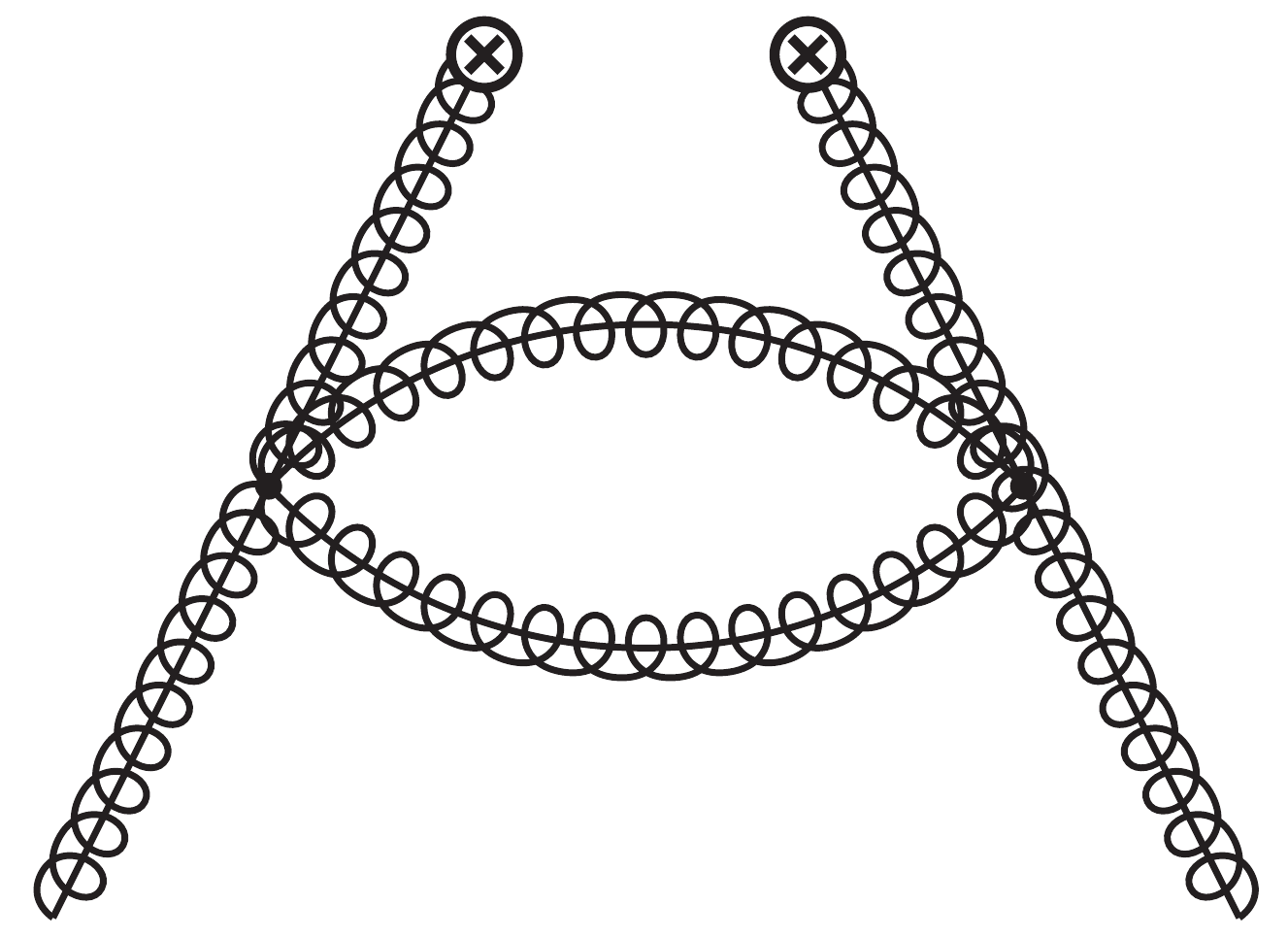}
\put(-100,68){e)}
\includegraphics[width=0.245\textwidth]{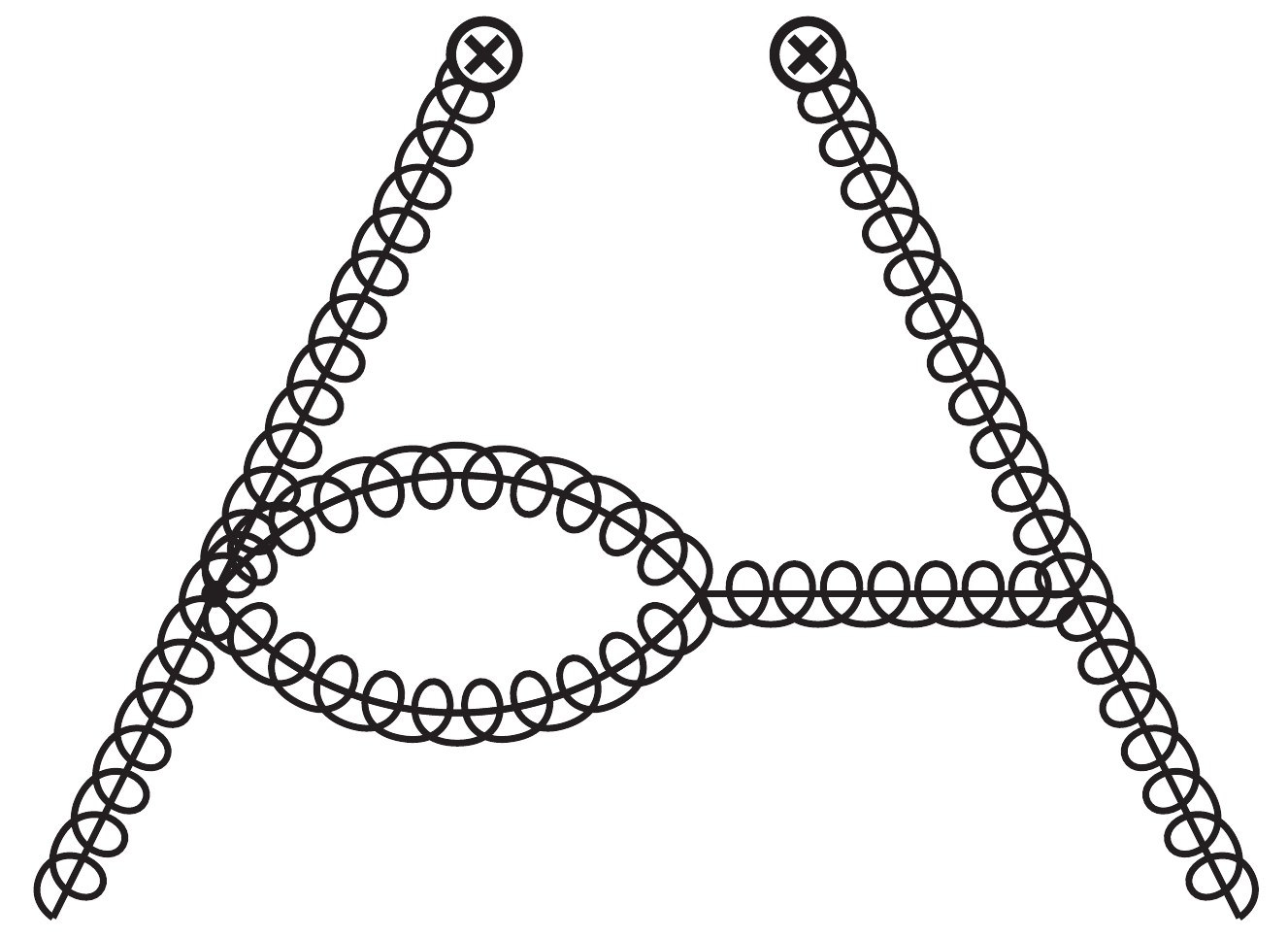}
\put(-100,68){f)}
\includegraphics[width=0.245\textwidth]{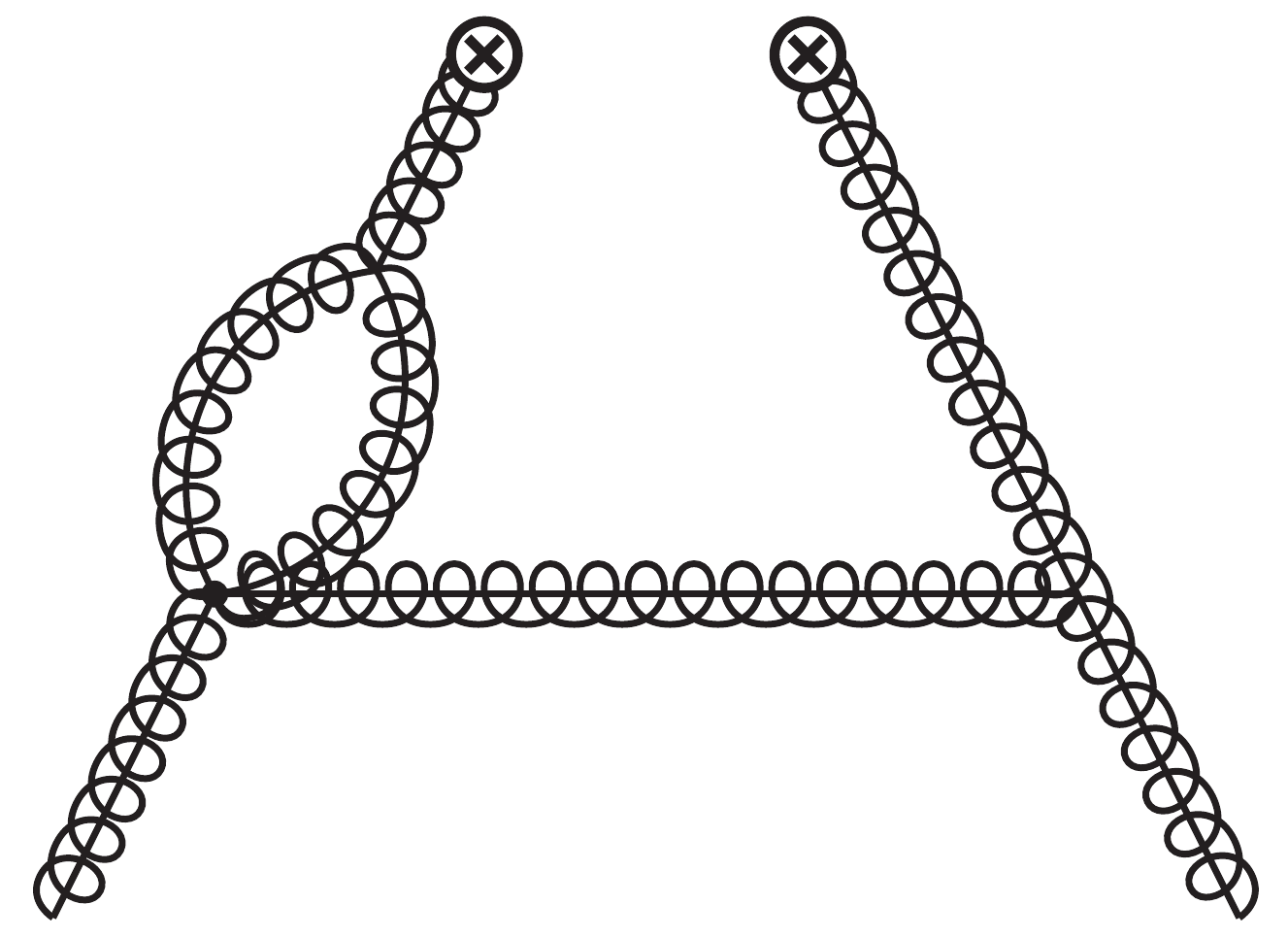}
\put(-100,68){g)}
\includegraphics[width=0.245\textwidth]{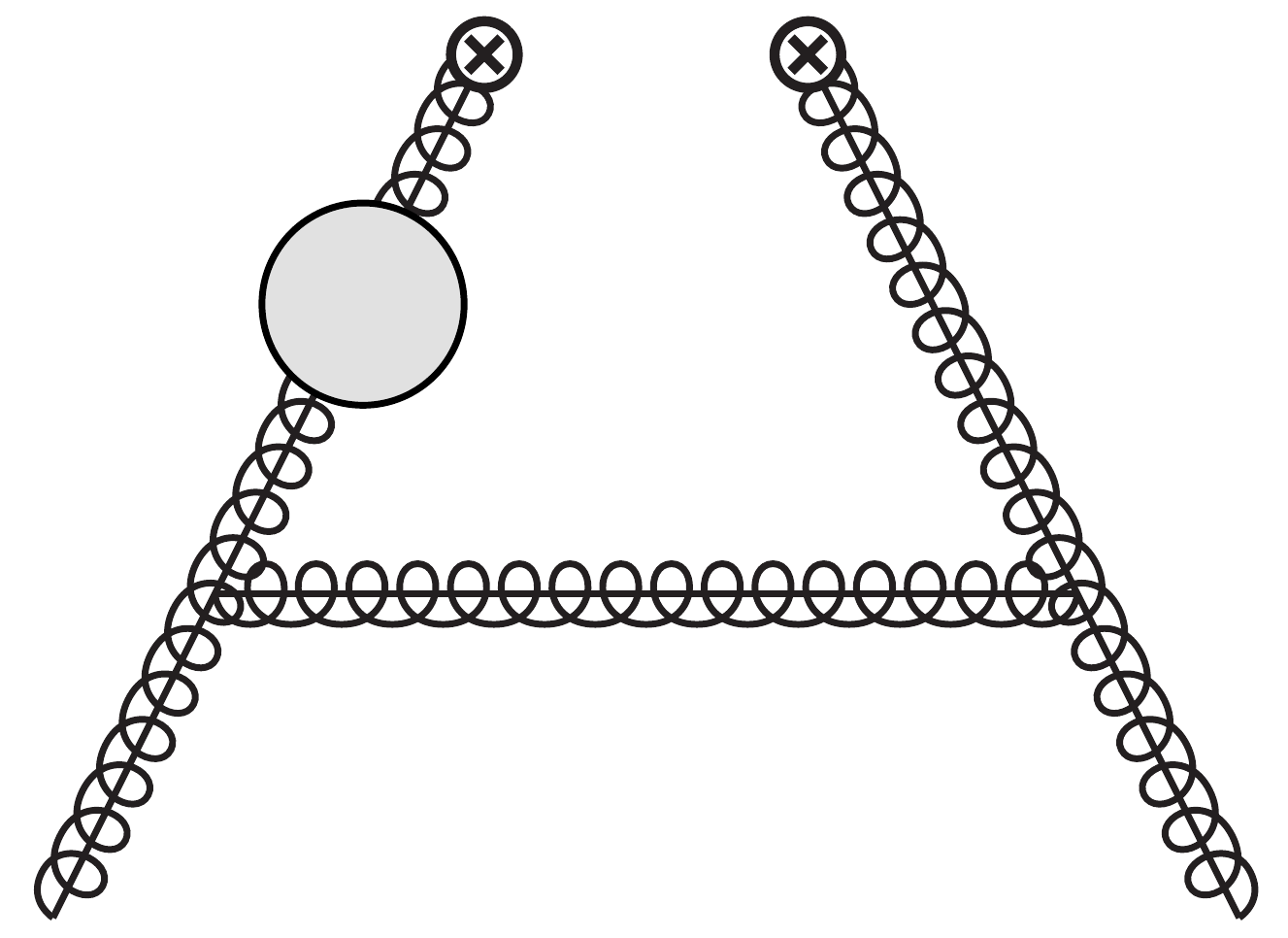}
\put(-100,68){h)}
\vspace{1ex}
\includegraphics[width=0.245\textwidth]{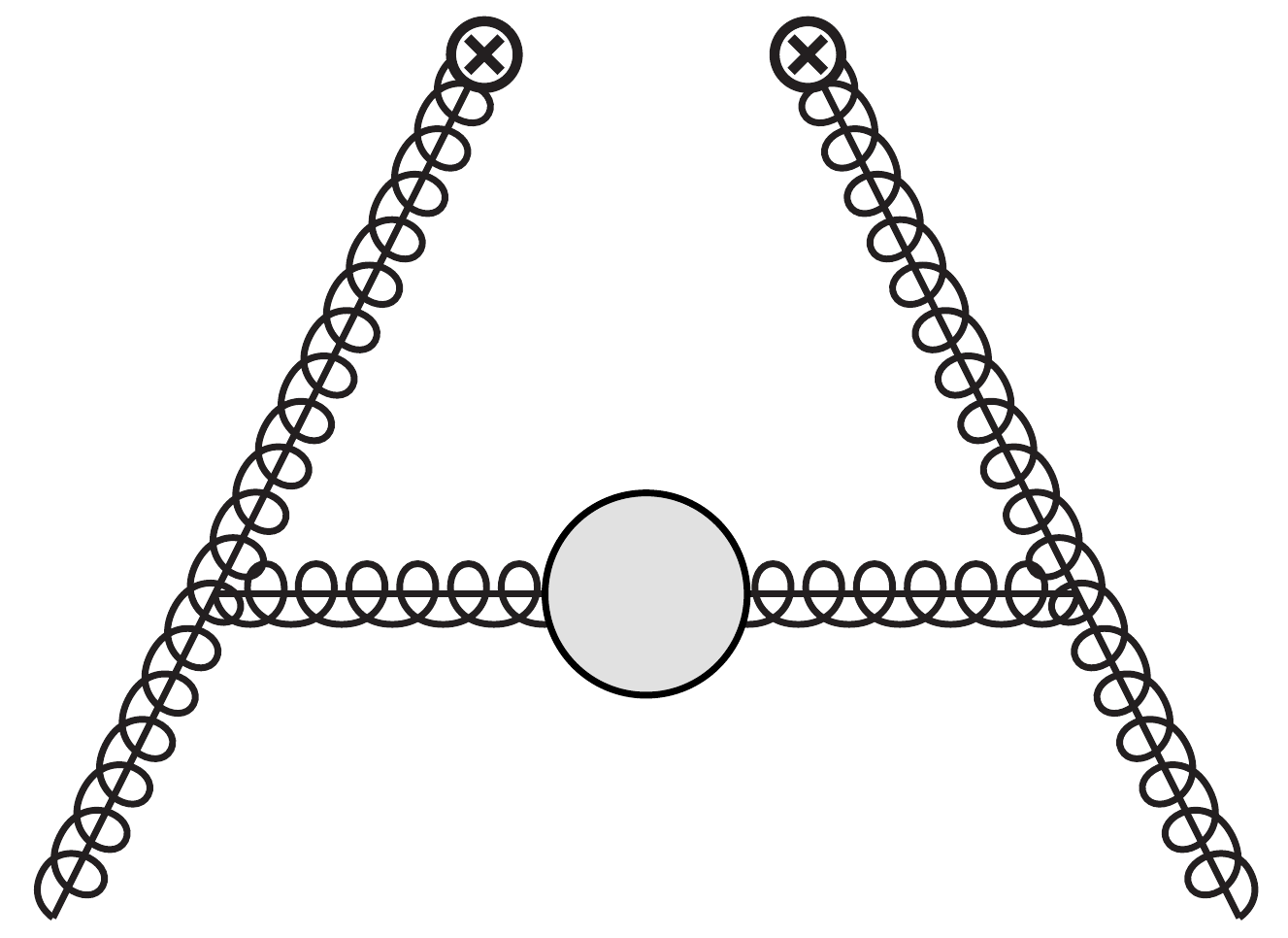}
\put(-100,68){i)}
\includegraphics[width=0.245\textwidth]{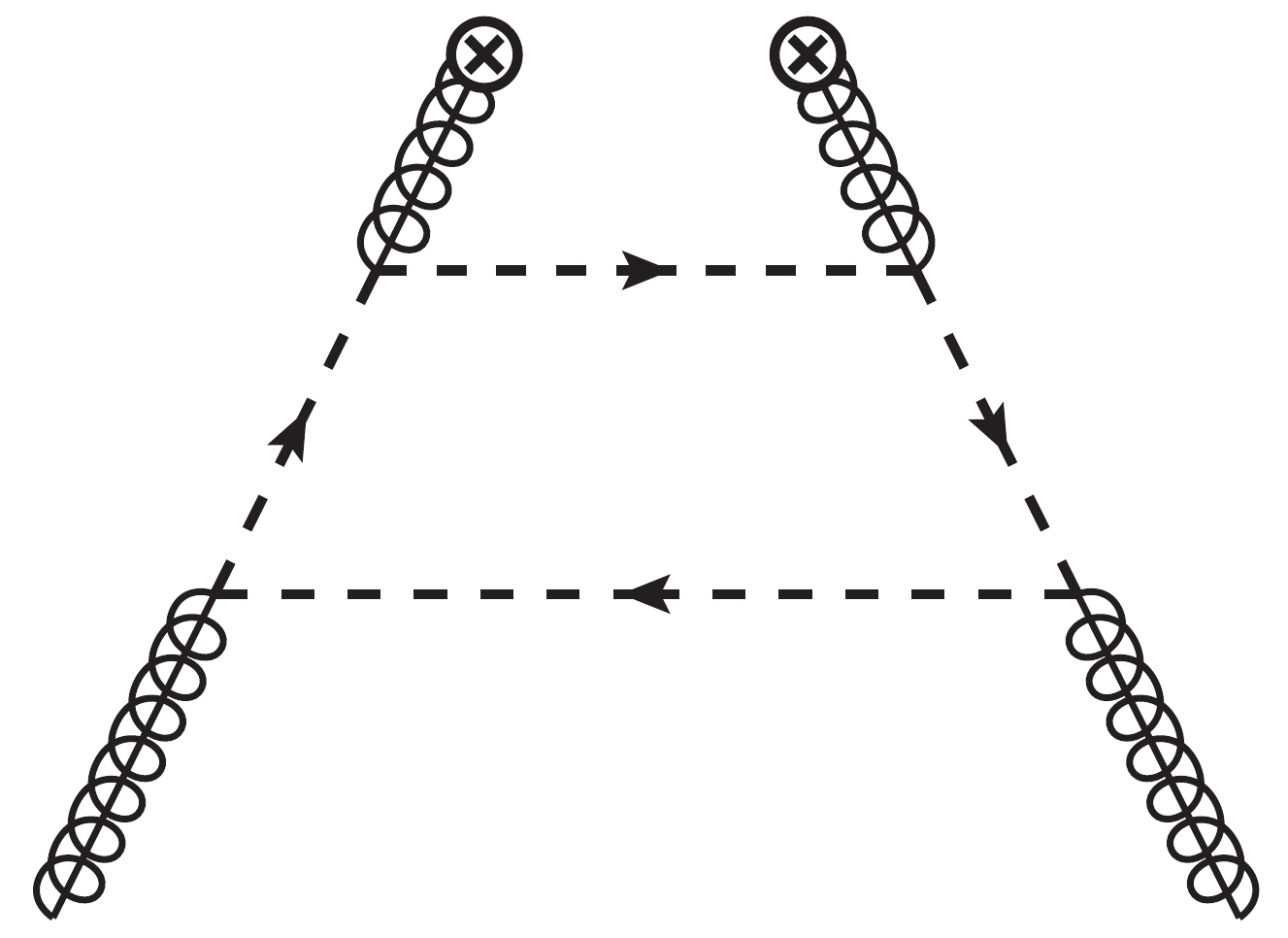}
\put(-100,68){j)}
\includegraphics[width=0.245\textwidth]{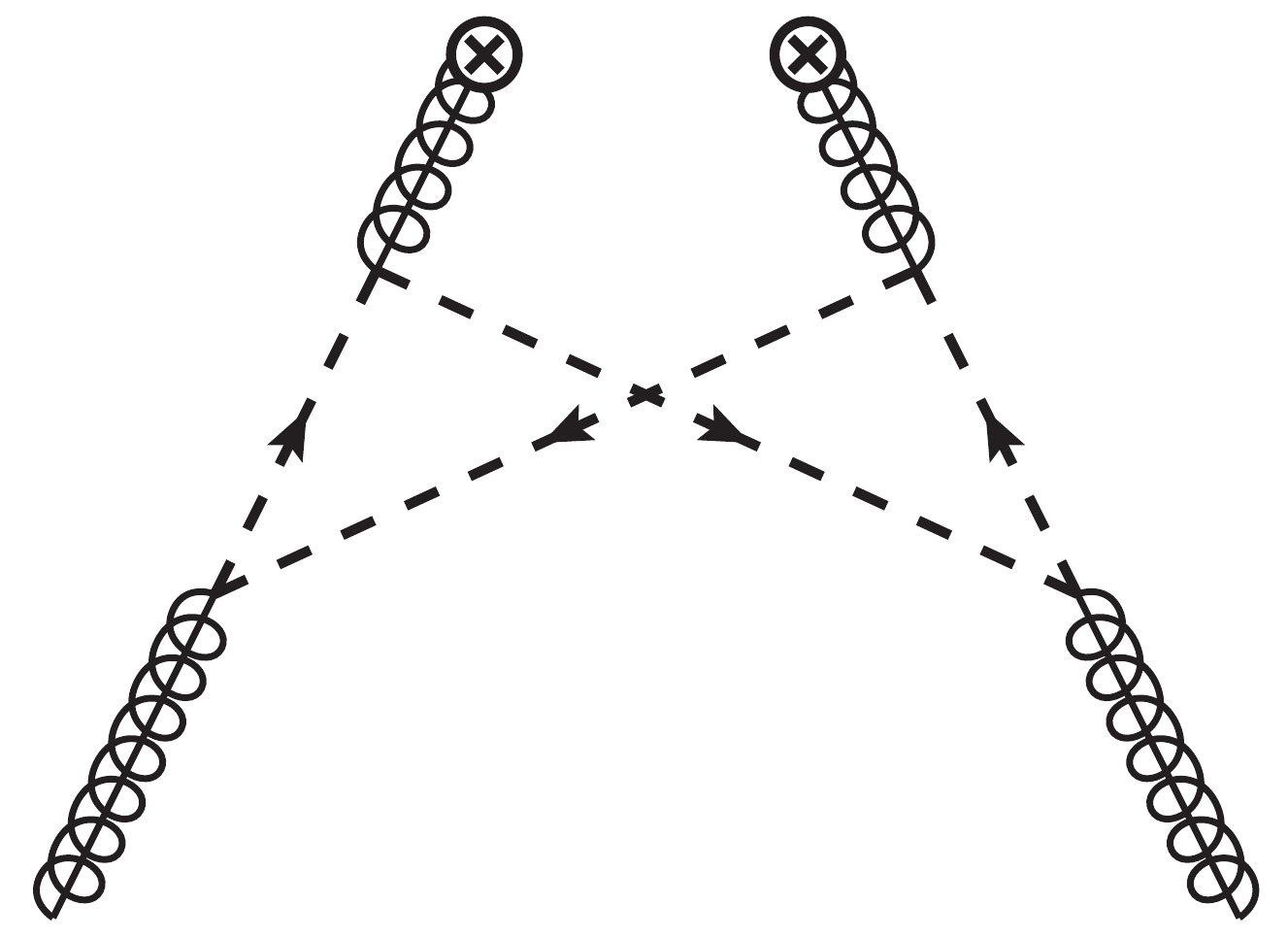}
\put(-100,68){k)}
\includegraphics[width=0.245\textwidth]{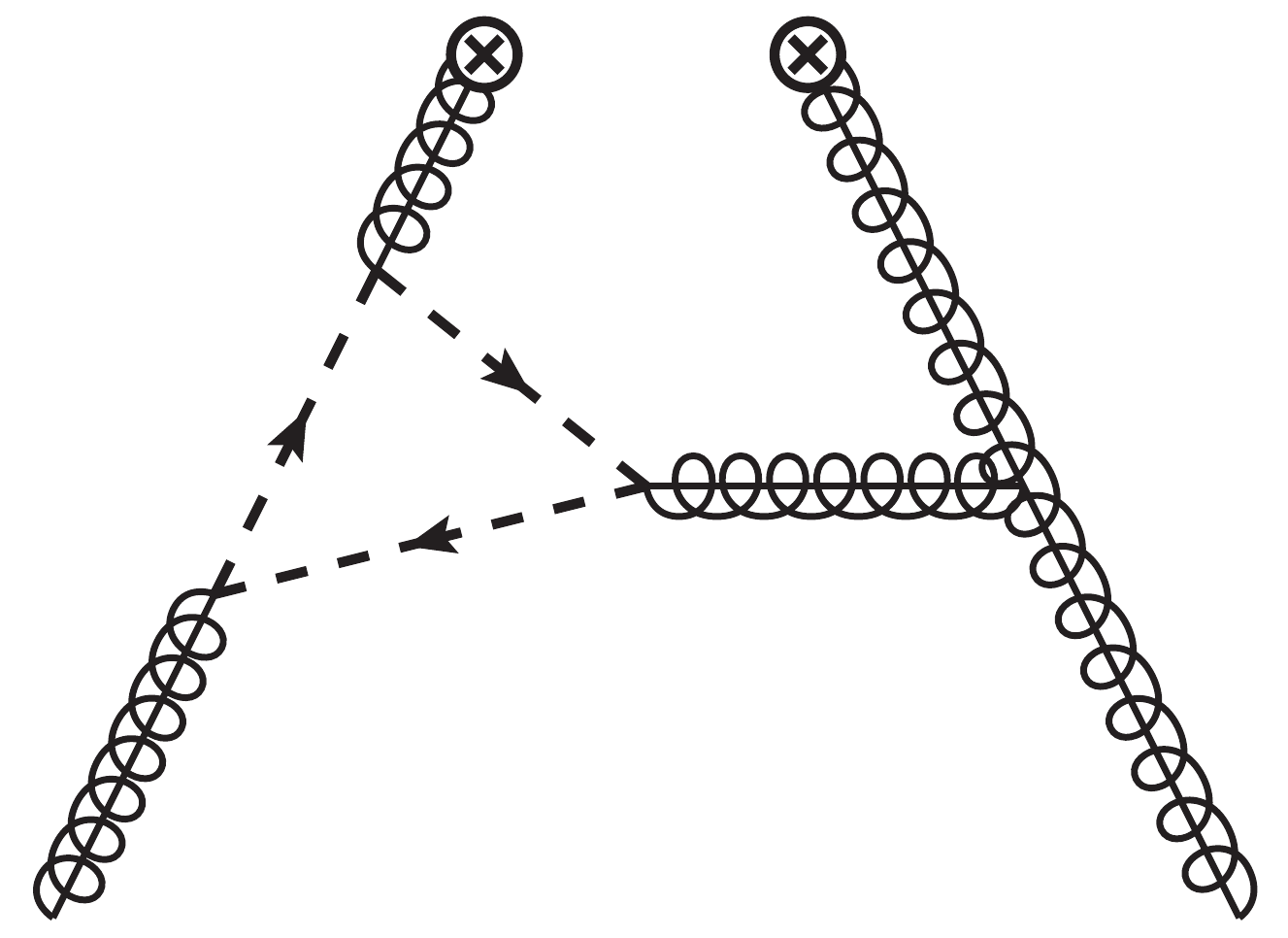}
\put(-100,68){l)}
\vspace{1ex}
\includegraphics[width=0.245\textwidth]{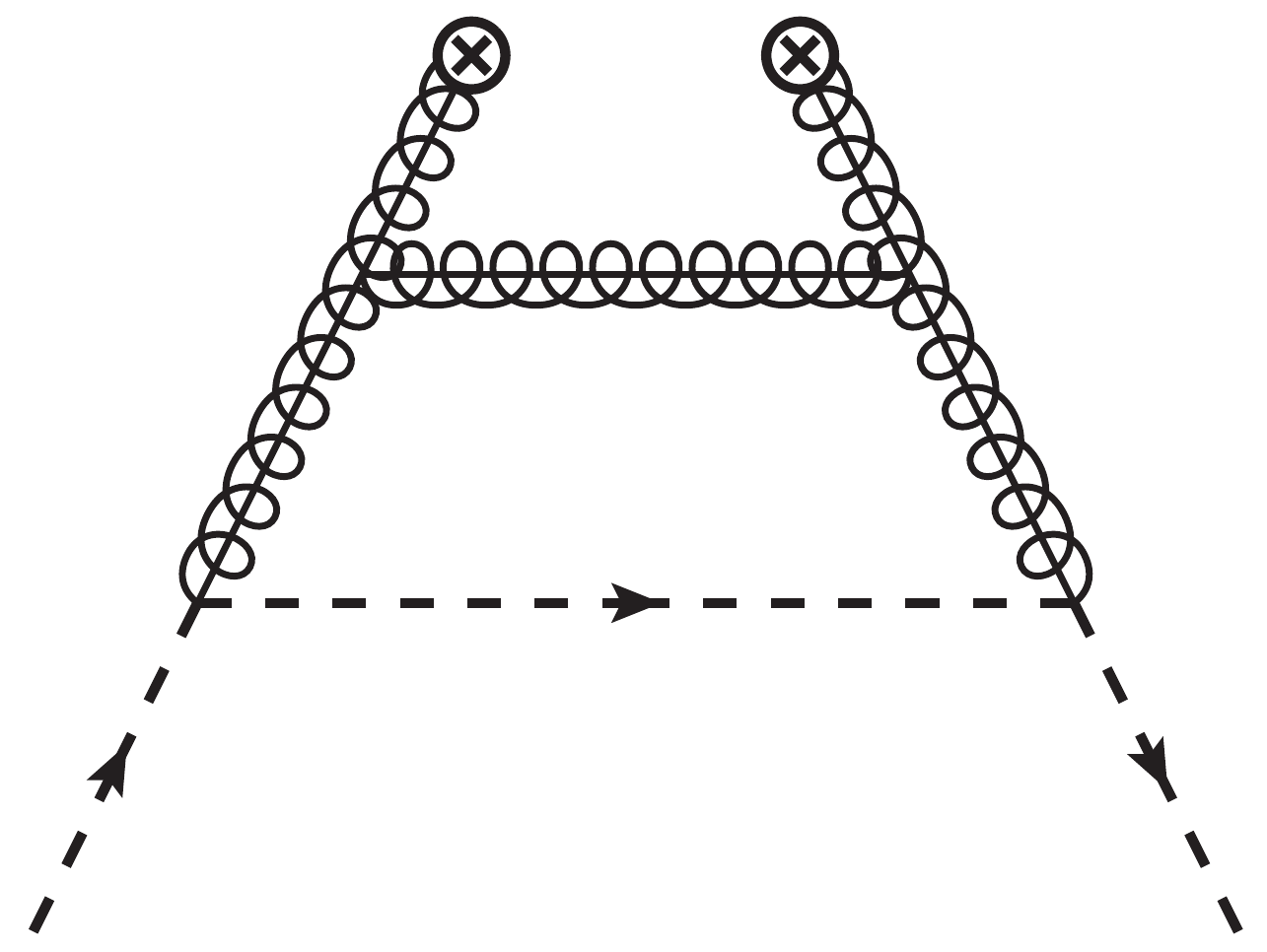}
\put(-100,68){m)}
\includegraphics[width=0.245\textwidth]{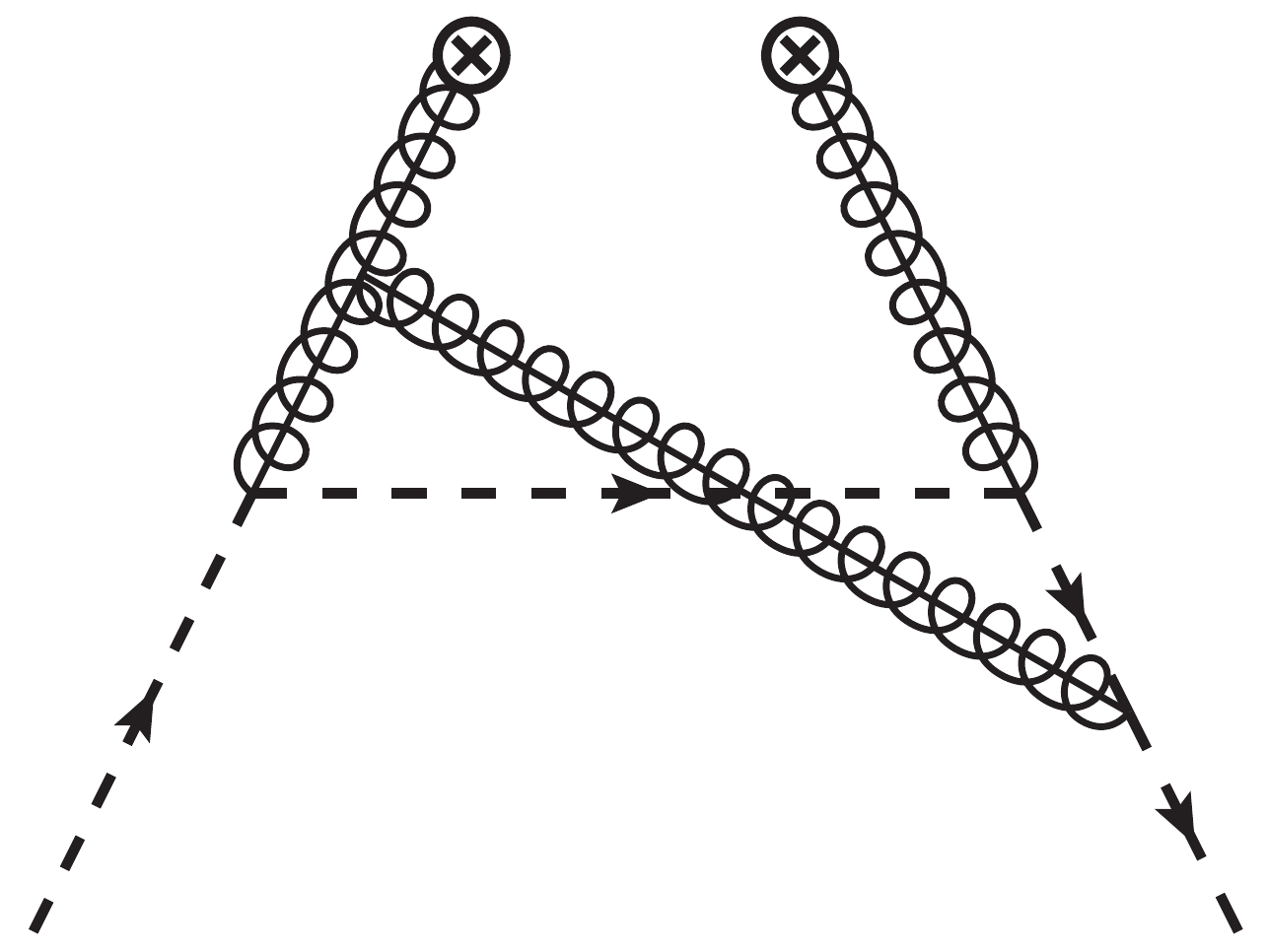}
\put(-100,68){n)}
\includegraphics[width=0.245\textwidth]{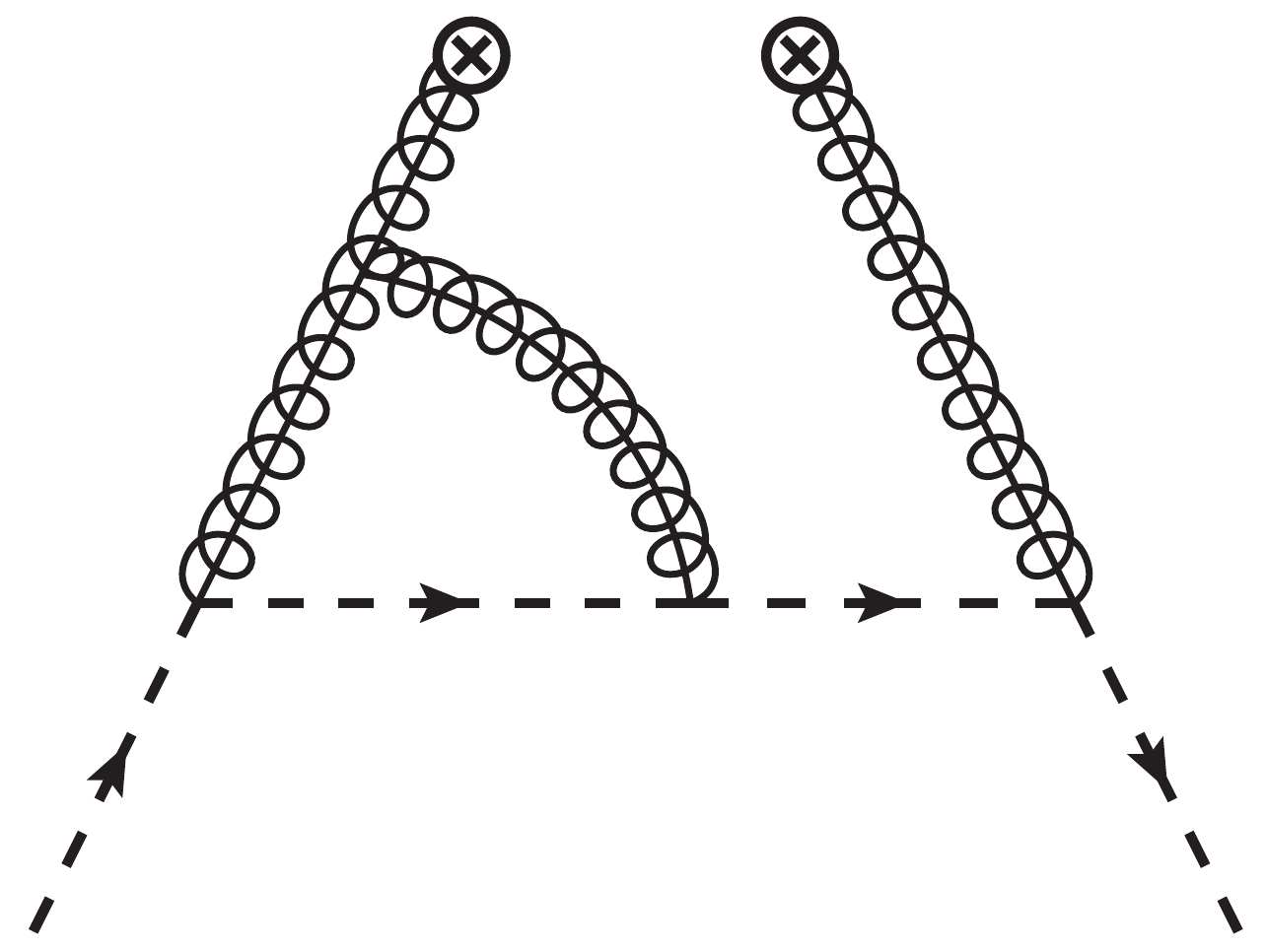}
\put(-100,68){o)}
\includegraphics[width=0.245\textwidth]{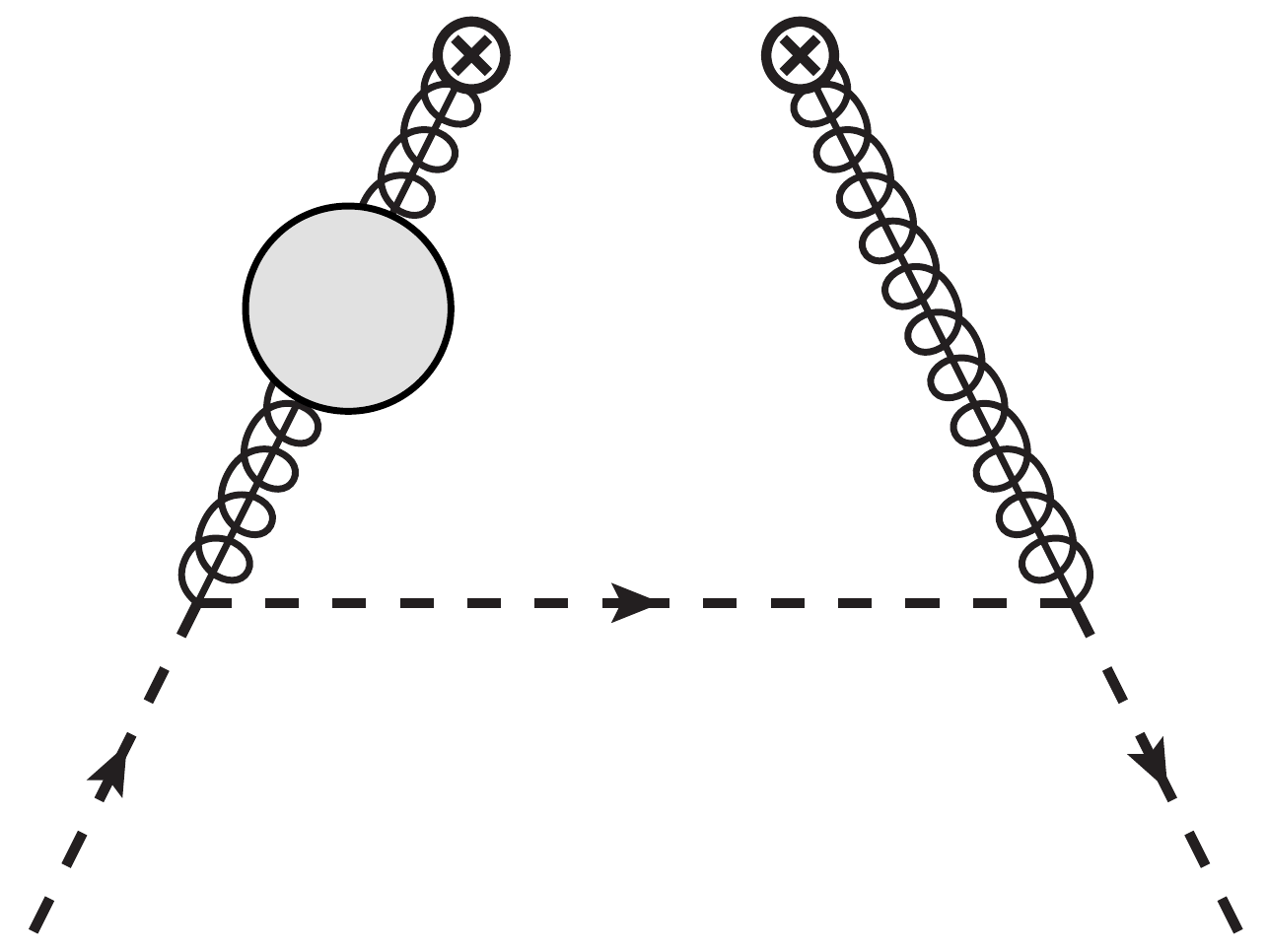}
\put(-100,68){p)}
\vspace{1ex}
\includegraphics[width=0.245\textwidth]{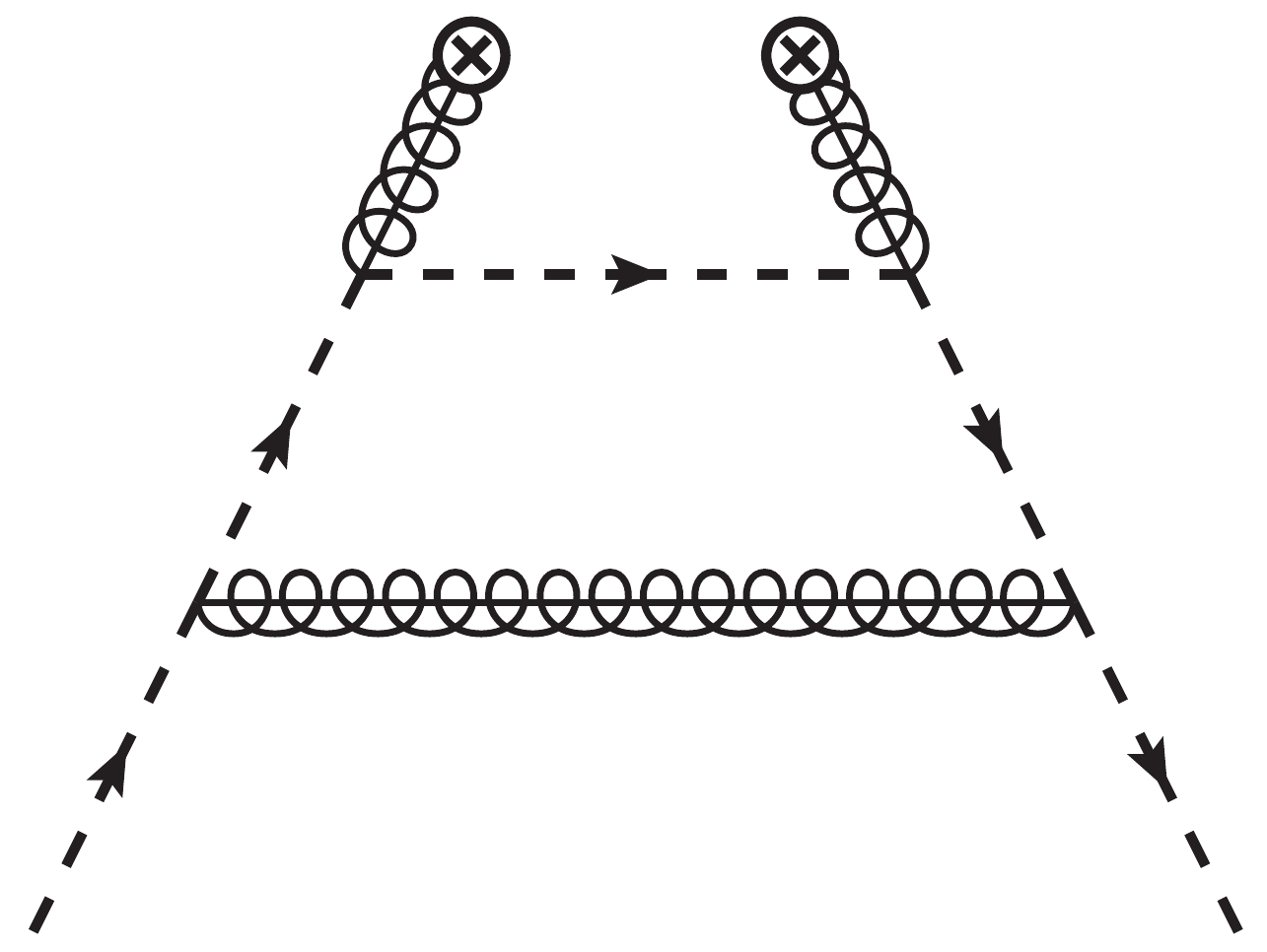}
\put(-100,68){q)}
\includegraphics[width=0.245\textwidth]{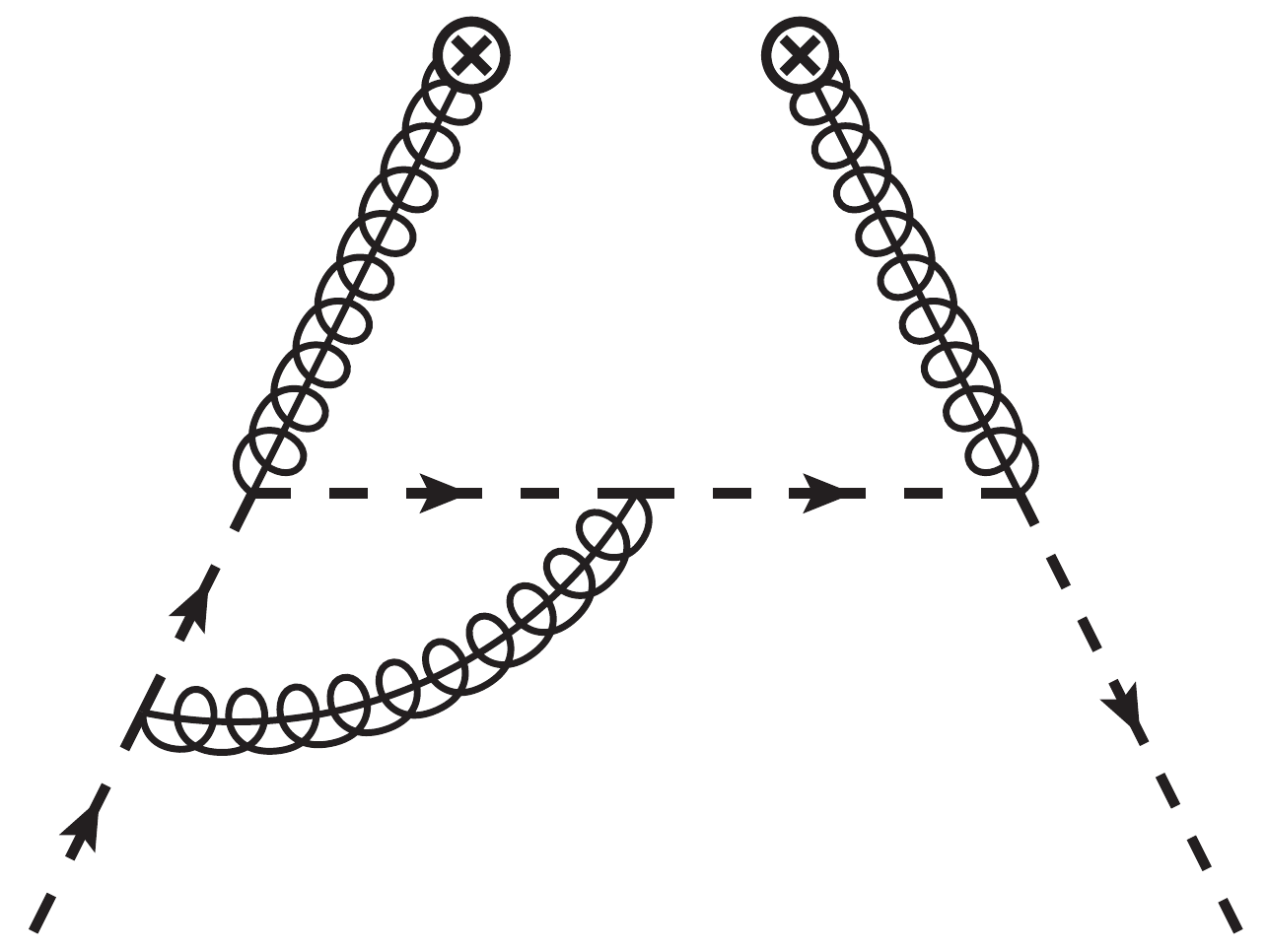}
\put(-100,68){r)}
\includegraphics[width=0.245\textwidth]{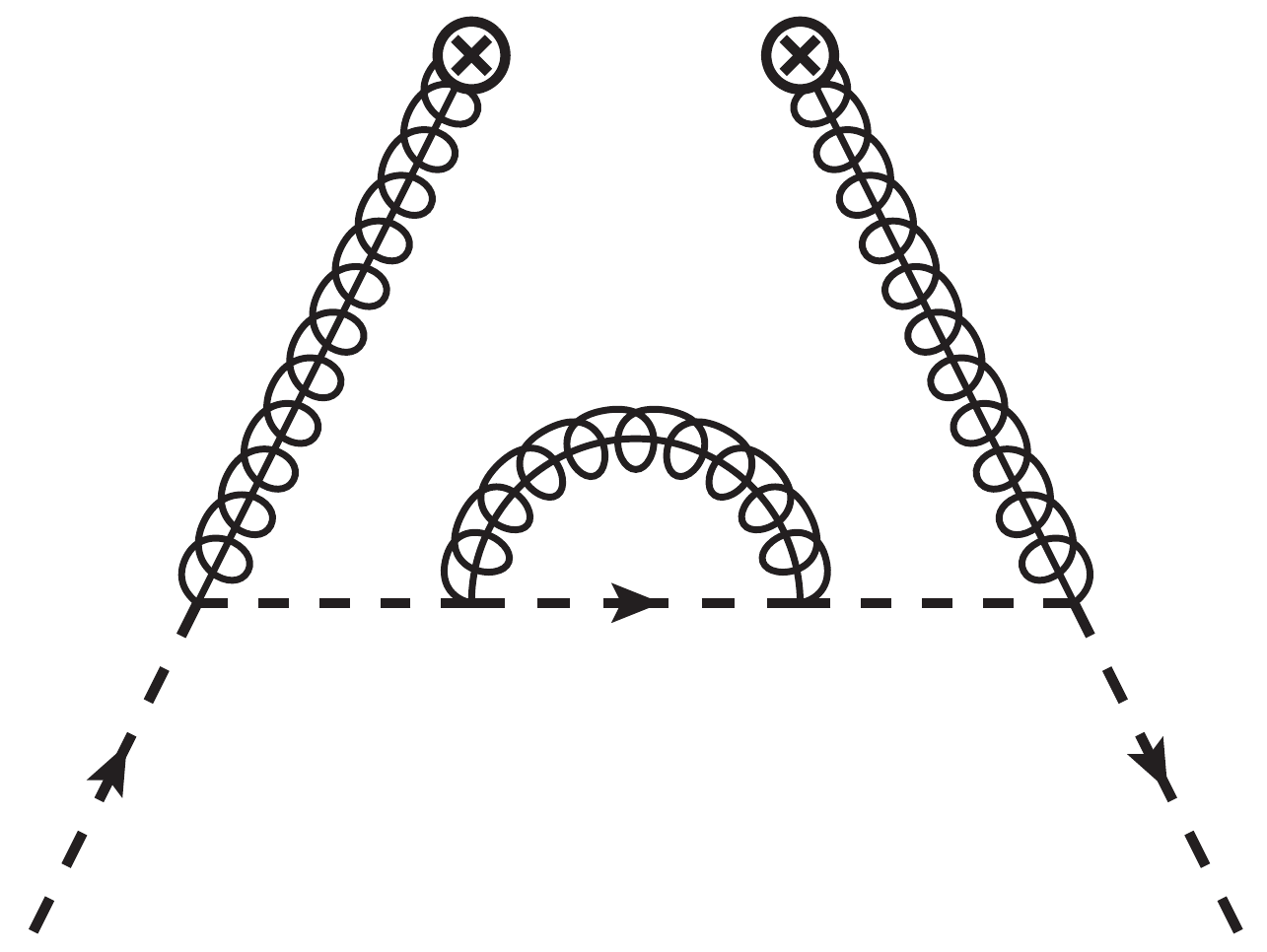}
\put(-100,68){s)}
\caption{
Diagrams contributing to the calculation of the NNLO matching coefficients $\cI_{gg}$ (a-l) and $\cI_{gq}$ (m-s) when using dimensional regularization. Left-right mirror graphs and graphs with reversed fermion flow in the loop are not displayed. 
The blob in diagrams~(h,i,p) represents the full one-loop gluon self-energy. 
The graphs can be computed using either standard QCD Feynman rules or SCET Feynman rules with collinear quark and gluon lines. Using axial gauge and QCD Feynman rules, this is the complete set of nontrivial diagrams. Using SCET Feynman rules, it has to be supplemented by diagrams involving vertices of four collinear particles. In Feynman gauge, additional diagrams with Wilson line connections (see e.g. \fig{endpoint}) or ghost loops contribute.
\label{fig:BgDiagrams}}
\end{figure}

We calculate the NNLO gluon beam function using the same methodology as in \mycite{Gaunt:2014xga}.
That is, we compute the bare (unrenormalized) partonic
beam function $B_{g/j}^{\bare}(t,z,g_0)$ from the discontinuity with respect to $t$ of the Feynman diagrams shown in
\fig{BgDiagrams}. From these we then extract the coefficients $\cI_{gj}(t,z,\mu)$ using the matching equations
given in section 2.2 of \mycite{Gaunt:2014xga}. We evaluate the diagrams
together with taking the discontinuity using two different methods, the `On-Shell Diagram
Method' and the `Dispersive Method', which are explained in \mycite{Gaunt:2014xga}.
We also perform the calculation using two different
gauges, namely light-cone axial ($\bar{n} \cdot A_n = 0$) gauge and Feynman gauge. The two different methods and gauge choices
gave the same final results, hence providing us with a strong cross check.

In the appendices, we provide some
further calculational details: in Appendix~\ref{sec:TChange}, we provide the
change of transverse variables in the On-Shell Diagram Method that is used to compute the cuts of 
the diagrams. In Appendix~\ref{sec:LCint}, we analyze the three-point integral that contains
the only (light-cone) divergence in the calculation that is not regulated by dimensional regularization 
(related to the discussion in section 3.1 of \mycite{Gaunt:2014xga}).
Finally in Appendix \ref{sec:W3vertex}, we complete the list of required SCET Feynman rules in Feynman gauge (see \mycite{Berger:2010xi}) by giving the expression for the triple gluon vertex associated with the gluon field strength operator $\cB_{n\perp}^\mu$, which is first needed at NNLO, and to our knowledge has not been given in the literature before.

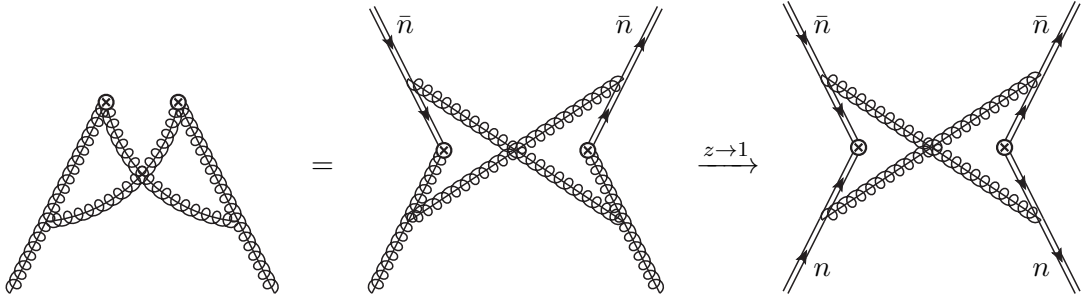
\begin{figure}[th]
\begin{center}
\includegraphics[width=0.25\textwidth]{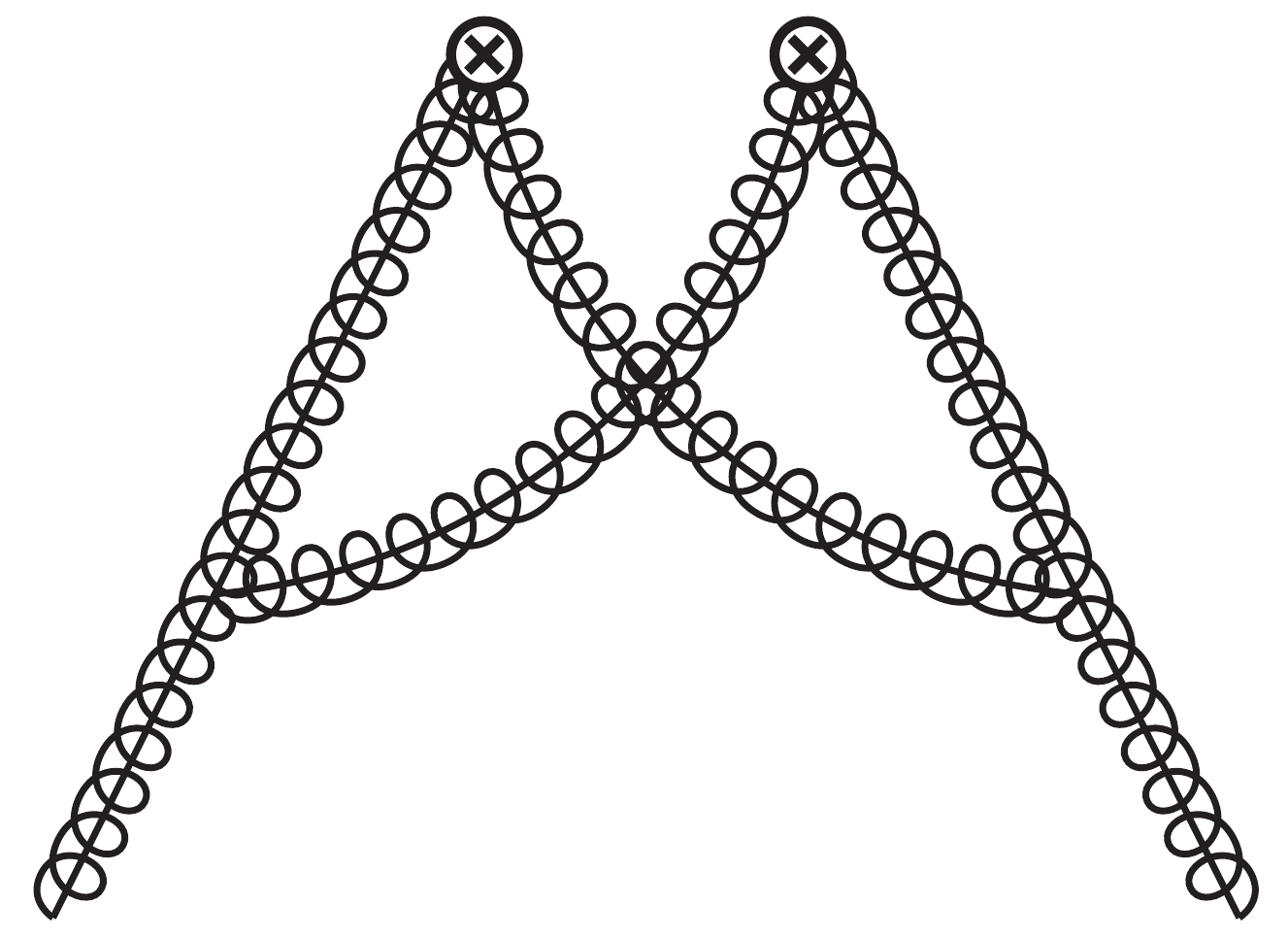} \raisebox{10 ex}{\;\;=\;\;}
\includegraphics[width=0.27\textwidth]{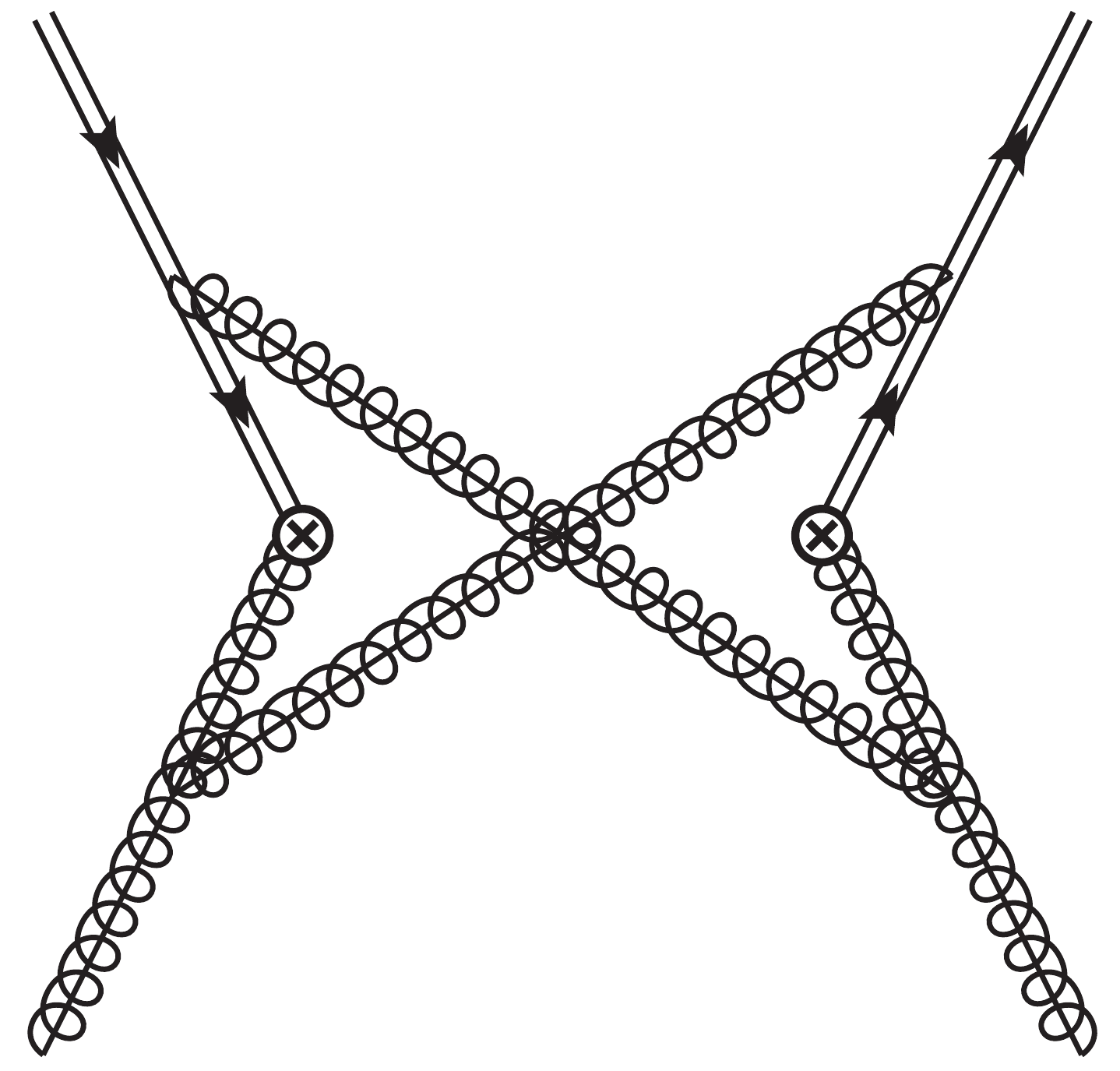} \raisebox{10 ex}{\;\;$\xrightarrow{z \to 1}$\;\;}
\put(-60,100){$\bn$}
\put(-142,100){$\bn$}
\includegraphics[width=0.27\textwidth]{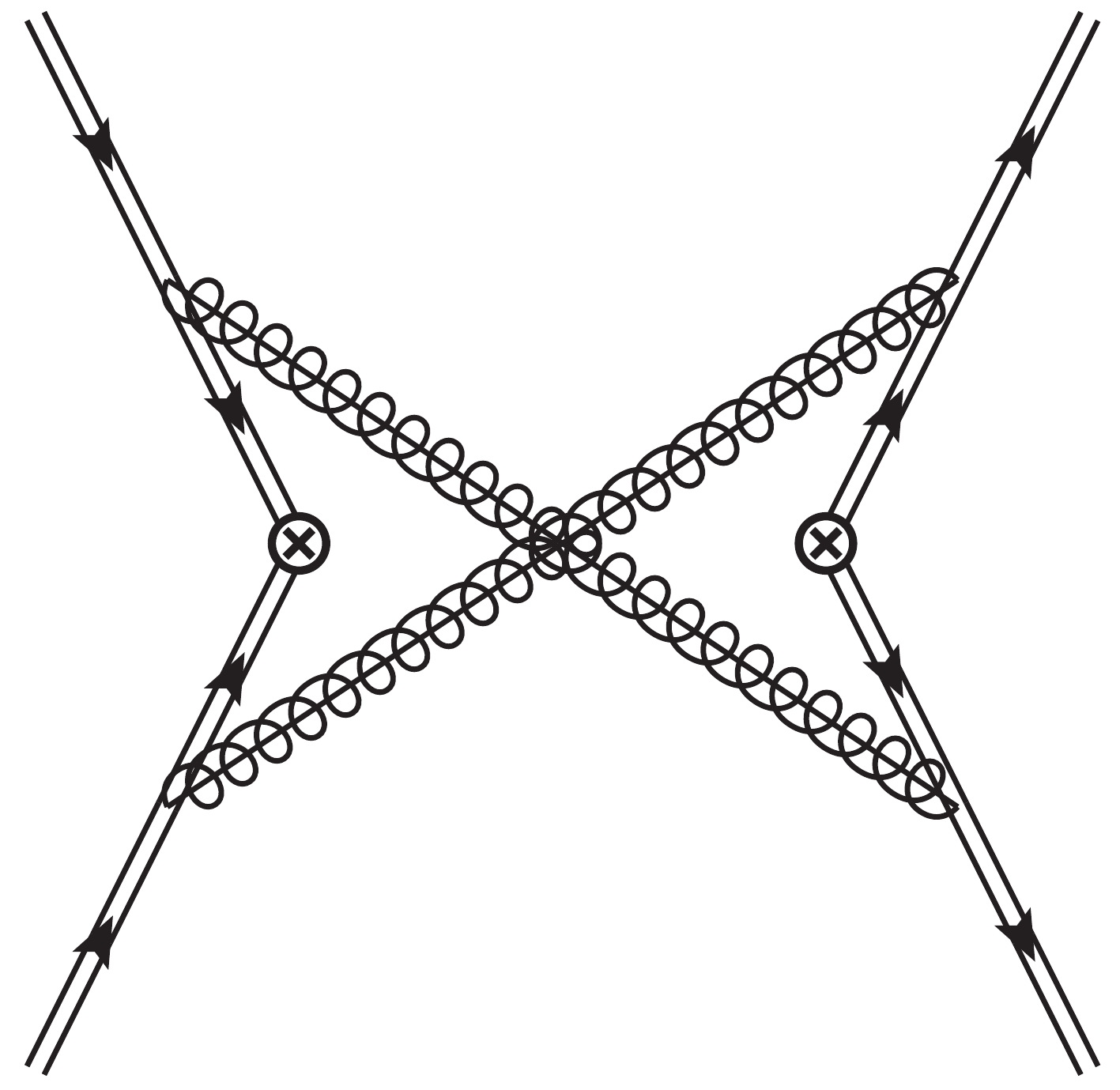}
\put(-20,100){$\bn$}
\put(-102,100){$\bn$}
\put(-20,8){$n$}
\put(-102,8){$n$}
\end{center}
\caption{
Diagrammatic calculation of the endpoint ($z\to1$) contributions to the partonic beam function in Feynman gauge.
The connections to the collinear Wilson lines in the beam function operator denoted as $\otimes$ in the left (example) diagram can also be drawn explicitly as connections to double lines that represent the Wilson lines along the $\bn$ direction in the adjoint representation (middle diagram). In the limit $z\to1$ the incoming gluon lines can be replaced by (adjoint) Wilson lines along the $n$ direction as shown in the right diagram. The calculation of the quark-quark channel endpoint is analogous, but with all the Wilson lines in the fundamental color representation.
\label{fig:endpoint}}
\end{figure}

To compute the endpoint ($z \to 1$) contributions to $B_{g/g}^{\bare}$, we deviate slightly from the procedure taken in~\cite{Gaunt:2014xga}.
In contrast to the quark case, we do not need to calculate the endpoint of $B_{g/g}^{\bare}$ directly. Rather, we appeal to the argument
given in \mycite{Gaunt:2014xga}, that the endpoint contributions can be obtained by replacing the incoming lines by (collinear)
Wilson lines, this time in the adjoint representation rather than the fundamental representation as was
appropriate for the quark case. This is illustrated in \fig{endpoint} for an example diagram with a nonzero endpoint contribution in Feynman gauge. Then the quark and gluon endpoint diagrams become identical, up to a
replacement of fundamental color matrices by adjoint color matrices for connections along the Wilson 
lines. At the two-loop order, this means that the quark and gluon endpoints
are actually equal, up to the replacement $C_F \to C_A$ in going from the quark to the gluon endpoint. Therefore,
we can obtain the gluon endpoint contribution simply by taking the quark endpoint contribution already calculated 
and making the replacement $C_F \to C_A$.\footnote{One can easily check that this works for the divergent terms $\propto \delta(1-z)$ in $B_{g/g}^{\bare}$, since they are fixed by the result for $z<1$.}

We regulate divergences associated with light-cone propagators
(in light-cone axial gauge) or Wilson line propagators (in Feynman gauge) by dimensional
regularization -- except for one particular case concerning diagram (c) that is discussed further in Appendix \ref{sec:LCint}. 
Other ways to consistently treat these light-cone divergences, employed in the past for the two-loop (axial gauge) calculations of the QCD splitting functions, are the principal value~\cite{Curci:1980uw, Furmanski:1980cm, Ellis:1996nn}, or the Mandelstam-Leibbrandt prescription~\cite{Mandelstam:1982cb, Leibbrandt:1983pj, Heinrich:1997kv, Bassetto:1998uv}.
Using dimensional regularization, we find in light-cone gauge that the swordfish diagrams (f) and (g) are zero,
whilst the diagram (d) contributes only a finite part, with no poles. Therefore our prescription gives
zero for the contributions of diagrams (d), (f), and (g) to the gluon-gluon splitting function, which is related to 
the infrared divergent part of the diagrams. In this respect, dimensional regularization behaves similarly to the principal value 
prescription, which in contrast to the Mandelstam-Leibbrandt prescription~\cite{Heinrich:1998mg} was observed to give zero contribution of the diagrams (d), (f), and (g) to the splitting function~\cite{Ellis:1996nn}.
In Feynman gauge, the diagrams (d), (f), and (g) contribute both to the poles and the finite pieces of $B_{g/g}^{\bare}$.

\section{Results}
\label{sec:results}

We expand the matching coefficient $\cI_{ij}$ in a perturbative series as
\begin{equation}
\cI_{ij} = \sum_{n=0}^{\infty} \biggl(\dfrac{\alpha_s}{4\pi}\biggr)^n \cI_{ij}^{(n)}
\,.\end{equation}
At tree level and one loop we have
\begin{align} \label{eq:I1master}
\cI_{ij}^\zero(t,z,\mu) &= \delta(t)\, \delta_{ij}\delta(1-z)
\nn\,,\\
\cI_{ij}^\one(t,z,\mu)
&= \frac{1}{\mu^2} \cL_1\Bigl(\frac{t}{\mu^2}\Bigr) \Gamma_0^i\, \delta_{ij}\delta(1 - z)
  + \frac{1}{\mu^2} \cL_0\Bigl(\frac{t}{\mu^2}\Bigr)
  \Bigl[- \frac{\gamma_{B\,0}^i}{2}\,\delta_{ij}\delta(1-z) + 2 P_{ij}^\zero(z) \Bigr]
  \nn \\ & \quad
  + \delta(t)\, 2I_{ij}^\one(z)
\,.\end{align}
Iteratively solving the renormalization group equation for $\cI_{ij}$ to $\ord{\alpha_s^2}$ yields for the two-loop matching coefficient~\cite{Gaunt:2014xga}
\begin{align} \label{eq:I2master}
\cI_{ij}^\two(t,z,\mu)
&= \frac{1}{\mu^2} \cL_3\Bigl(\frac{t}{\mu^2}\Bigr) \frac{(\Gamma_0^i)^2}{2}\, \delta_{ij}\delta(1-z)
  \nn \\ & \quad
  + \frac{1}{\mu^2} \cL_2\Bigl(\frac{t}{\mu^2}\Bigr)
  \Gamma_0^i \Bigl[- \Bigl(\frac{3}{4} \gamma_{B\,0}^i + \frac{\beta_0}{2} \Bigr) \delta_{ij}\delta(1-z) + 3P^\zero_{ij}(z) \Bigr]
  \nn \\ & \quad
  + \frac{1}{\mu^2} \cL_1\Bigl(\frac{t}{\mu^2}\Bigr)
  \biggl\{ \Bigl[\Gamma_1^i - (\Gamma_0^i)^2 \frac{\pi^2}{6} + \frac{(\gamma_{B\,0}^i)^2}{4} + \frac{\beta_0}{2} \gamma_{B\,0}^i \Bigr] \delta_{ij}\delta(1-z)
  \nn \\ & \qquad
  + 2\Gamma_0^i\, I^\one_{ij}(z)
  - 2 (\gamma_{B\,0}^i + \beta_0) P^\zero_{ij}(z)
  + 4 \sum_k P^\zero_{ik}(z)\conv_z P^\zero_{kj}(z) \biggr\}
  \nn \\ & \quad
  + \frac{1}{\mu^2} \cL_0\Bigl(\frac{t}{\mu^2}\Bigr)
  \biggl\{ \Bigl[(\Gamma_0^i)^2 \zeta_3 + \Gamma_0^i \gamma_{B\,0}^i \frac{\pi^2}{12} - \frac{\gamma_{B\,1}^i}{2} \Bigr]
  \delta_{ij} \delta(1-z)
  - \Gamma_0^i \frac{\pi^2}{3} P^\zero_{ij}(z)
  \nn \\ & \qquad
  - (\gamma_{B\,0}^i + 2\beta_0) I^\one_{ij}(z)
  + 4 \sum_k I^\one_{ik}(z)\conv_z P^\zero_{kj}(z)
  + 4 P^\one_{ij}(z)
  \biggr\}
  \nn \\ & \quad
  + \delta(t)\, 4 I^\two_{ij}(z)
\,,\end{align}
where $\beta_0 = (11 C_A - 4 T_F n_f)/3$, and
\begin{align} \label{eq:plusdefmaintxt}
\cL_n(x)
&= \biggl[ \frac{\theta(x) \ln^n x}{x}\biggr]_+
 = \lim_{\eps \to 0} \frac{\df}{\df x}\biggl[ \theta(x- \eps)\frac{\ln^{n+1} x}{n+1} \biggr]
\end{align}
denotes the usual plus distributions. The new results of our calculation are the two-loop gluon $\delta(t)$-terms $I^\two_{gj}(z)$ in the last line of \eq{I2master}. All remaining ingredients in \eq{I2master} are known and, for the case $i=g$, have been given in \mycite{Berger:2010xi}. They are collected in \app{pert} for completeness.

We write the $I^\two_{gj}(z)$ as
\begin{align}
I_{gg}^\two(z) &= \theta(z) \bigl[ C_A I_{ggA}^\two(z) + T_F n_f I_{ggF}^\two(z) \bigr]
\,, \nn \\
I_{g q_i}^\two(z) = I_{g \bar q_i}^\two(z) &= C_F\, \theta(z) I_{gq}^\two(z)
\,,\end{align}
and we find
\begin{align} \label{eq:Igg_two}
I_{ggA}^\two(z)
&= \delta(1-z) \Bigl[ C_A \Bigl( \frac{52}{27} - \frac{\pi^2}{6} + \frac{11\pi^4}{360} \Bigr)
      + \beta_0 \Bigl( \frac{41}{27} - \frac{5 \pi^2}{24} - \frac{5 \zeta_3}{6} \Bigr) \Bigr]
   \nn \\ & \quad
   + C_A \biggl\{
      \frac{2(1 - z + z^2)^2}{z} \Bigl[\cL_3(1-z) + \Bigl(\frac{2}{3} - \pi^2\Bigr) \cL_1(1-z)
      + \Bigl(-\frac{8}{9} + \frac{15 \zeta_3}{2}\Bigr)\cL_0(1-z) \Bigr]
      \nn \\ & \qquad
      + P_{gg}(z) \bigl[V_3(z) - U_3(z) - T_3(z) - \ln\frac{1-z}{z} \ln(1-z) \ln z \bigr]
      + 8(1 + z) T_3(z)
      \nn \\ & \qquad
      + P_{gg}(-z) S_3(z)
      + \Bigl(- \frac{22}{3 z} + 6- 6z + \frac{22 z^2}{3}\Bigr) \Bigl[\ln^2\frac{1 - z}{z} - \frac{\pi^2}{3} \Bigr]
      \nn \\ & \qquad
      + \Bigl(\frac{11}{3 z} + 15 + 4 z + \frac{44 z^2}{3}\Bigr) \ln^2z
      + \Bigl(\frac{22}{3 z} + 14 + 8 z + \frac{44 z^2}{3}\Bigr) \Bigl[\Li_2(z) - \frac{\pi^2}{6} \Bigr]
      \nn \\ & \qquad
      + \Bigl(- \frac{143}{18 z} + \frac{34}{3} - \frac{145 z}{12} + \frac{143 z^2}{18}\Bigr) \ln(1-z)
      \nn \\ & \qquad
      - \Bigl[\frac{4}{3 (1 - z)} + \frac{149}{18 z} + \frac{155}{6} + \frac{101 z}{12} + \frac{43 z^2}{2} \Bigr] \ln z
      -(1-z) \Bigl( \frac{209}{9 z} + 8 + \frac{403 z}{18} \Bigr)
   \biggr\}
   \nn \\ & \quad
   + \beta_0 \biggl\{
      \frac{2(1 - z + z^2)^2}{z}\Bigl[-\frac{1}{4} \cL_2(1-z) + \frac{5}{6} \cL_1(1-z) + \Bigl(-\frac{7}{9} + \frac{\pi^2}{12}\Bigr) \cL_0(1-z) \Bigr]
      \nn \\ & \qquad
      + P_{gg}(z) \Bigl[ \frac{1}{2} \ln(1-z)\ln z - \frac{1}{4} \ln^2 z - \frac{5}{6} \ln z \Bigr]
      + \Bigl(\frac{13}{6 z} - \frac{3}{2} + \frac{7 z}{4} - \frac{13 z^2}{6} \Bigr) \ln\frac{1-z}{z}
      \nn \\ & \qquad
      - (1 + z) \Bigl[\Li_2(z) + \frac{3}{4} \ln^2 z + \frac{7}{12} \ln z - \frac{\pi^2}{6} \Bigr]
      - \frac{34}{9 z} + \frac{19}{6} - \frac{11 z}{3} + \frac{77 z^2}{18}
   \biggr\}
\,, \nn \\
I_{ggF}^\two(z)
&= C_F \biggl\{
      - 4(1 + z) T_3(z)
      + \Bigl(\frac{4}{3 z} + 1 - z - \frac{4 z^2}{3}\Bigr) \Bigl[\ln^2\frac{1 - z}{z} - \frac{\pi^2}{3} \Bigr]
      + \frac{5+7z}{2} \ln^2z
      \nn \\ & \qquad
      + 2(2+3z) \Bigl[\Li_2(z) + \ln z - \frac{\pi^2}{6} \Bigr]
      + \Bigl(- \frac{14}{9 z} - \frac{40}{3} + \frac{28 z}{3} + \frac{50 z^2}{9}\Bigr) \ln\frac{1-z}{z}
      \nn \\ & \qquad
      + \frac{23}{27 z} + \frac{247}{9} - \frac{211 z}{9} - \frac{131 z^2}{27}
   \biggr\}
\,,\end{align}
and
\begin{align} \label{eq:Igq_two}
I_{gq}^\two(z)
&= C_A \biggl\{
      P_{gq}(z) \Bigl[V_3(z) - U_3(z) - T_3(z) + \frac{5}{6} \ln^3(1-z)
         - \frac{2\pi^2}{3} \ln(1-z) - \frac{\pi^2}{3}  \ln z
      \nn \\ & \qquad
         + \frac{45}{6} \zeta_3 \Bigr]
      + 2(4+z) T_3(z)
      + P_{gq}(-z)\,  S_3(z)
      - \frac{z}{2}\, \Bigl[S_2(z) - \frac{\pi^2}{2} \Bigr]
      \nn \\ & \qquad
      + \Bigl(-\frac{31}{6z} + 4 + 2 z + \frac{2 z^2}{3} \Bigr) \Bigl[\ln^2\frac{1 - z}{z} - \frac{\pi^2}{3} \Bigr]
      + \Bigl(\frac{11}{3 z} + 15 + \frac{11 z}{4} + \frac{8 z^2}{3} \Bigr) \ln^2z
      \nn \\ & \qquad
      + \Bigl(\frac{22}{3 z} + 14 + 5 z + \frac{8 z^2}{3} \Bigr) \Bigl[\Li_2(z) - \frac{\pi^2}{6} \Bigr]
      + \Bigl(- \frac{181}{18 z} + \frac{37}{3} - 6 z + \frac{44 z^2}{9} \Bigr) \ln(1-z)
      \nn \\ & \qquad
      - \Bigl(\frac{43}{6 z} + \frac{51}{2} + \frac{13 z}{6} + \frac{88 z^2}{9} \Bigr) \ln z
      - \frac{2351}{108 z} + \frac{101}{6} - \frac{83 z}{36} + \frac{152 z^2}{27}
   \biggr\}
   \nn \\ & \quad
   + C_F \biggl\{
      P_{gq}(z) \Bigl[\frac{1}{6} \ln^3(1-z) - \ln\frac{1-z}{z} \ln(1-z) \ln z - \frac{\pi^2}{3} \ln\frac{1-z}{z} \Bigr]
      - (2 - z)\, T_3(z)
      \nn \\ & \qquad
      + \Bigl(- \frac{9}{2 z} + \frac{11}{2} - 3 z \Bigr) \Bigl[\ln^2\frac{1 - z}{z} - \frac{\pi^2}{3} \Bigr]
      + \Bigl(- \frac{3}{z} + 3 - \frac{5 z}{2} \Bigr) \Bigl[\ln(1-z) \ln z + \frac{\pi^2}{6}\Bigr]
      \nn \\ & \qquad
      + \Bigl(\frac{9}{2 z} - 7 + \frac{9 z}{8}\Bigr) \ln^2z
      - \frac{6 + 5 z}{2}\, \Bigl[\Li_2(z) - \frac{\pi^2}{6} \Bigr]
      + \Bigl(\frac{21}{2z} - \frac{75}{6} \Bigr) \ln(1-z)
      \nn \\ & \qquad
      + \Bigl(-\frac{15}{6 z} + \frac{9}{4} + \frac{15z}{4}\Bigr) \ln z
      - \frac{43}{4z} + 16  - 3 z
   \biggr\}
   \nn \\ & \quad
   + \beta_0 \biggl\{
      P_{gq}(z) \Bigl[\frac{1}{4} \ln^2(1-z) - \frac{1}{2} \ln^2z + \frac{5}{6}\ln(1-z) -\frac{5}{3} \ln z - \frac{14}{9} \Bigr]
      \nn \\ & \qquad
      + \frac{z}{2} \Bigl[\ln(1-z) + \frac{5}{3} \Bigr]
   \biggr\}
\,.\end{align}
For simplicity we have suppressed the overall $\theta(1-z)$ multiplying the regular terms. The auxiliary functions,
\begin{align} \label{eq:Sdef}
S_2(z) &= -2 \Li_2(-z) - 2 \ln(1+z)\ln z - \frac{\pi^2}{6}
\,, \nn \\
S_3(z) 
&= 2 \Li_3(1 - z) - \Li_3(z) + 4 \Li_3\Bigl(\frac{1}{1 + z}\Bigr) - \Li_3(1 - z^2)
   + \frac{\pi^2}{3} \ln(1 + z) - \frac{2}{3} \ln^3(1 + z)
   \nn \\ & \quad
   - \frac{5\zeta_3}{2} + \frac{\pi^2}{6} \ln z + S_2(z) \ln\frac{1-z}{z} - \frac{\ln^3z}{4}
\,, \nn \\
T_3(z)
&=  \Li_3(1 - z) - \Li_2(1 - z)\,\ln(1 - z)
   - \Bigl[\Li_2(z) + \frac{1}{2} \ln^2(1 - z) + \frac{5}{12} \ln^2z - \frac{\pi^2}{3} \Bigr] \ln z
\,, \nn \\
U_3(z)
&= -4 \Li_3(1 - z) + \Li_3(z) - \zeta_3 - \ln(1 - z)\Bigl[\Li_2(z) - \frac{\pi^2}{6} \Bigr] + 2 \Li_2(1 - z)\, \ln z - \frac{\ln^3z}{4}
\,, \nn \\
V_3(z)
&= -4 \Li_3(1 - z) - 5 \Li_3(z) + 5 \zeta_3 + \frac{1}{2} \ln(1 - z) \ln^2z
   \nn \\ & \quad
   - \Bigl[2 \ln^2(1 - z) + \frac{11}{12} \ln^2z -\frac{13\pi^2}{6} \Bigr] \ln z
\,,\end{align}
all vanish for $z\to 1$ at least like $1-z$ and are identical to those for the quark case~\cite{Gaunt:2014xga}.

As in the quark beam function calculation, one can extract from the poles of the bare beam function either
the two-loop anomalous dimension for the gluon beam function $\gamma_{B1}^g(t,\mu)$ , or the two-loop splitting 
functions $P_{gi}^{(1)}(z)$, assuming the other quantity is known (alternatively one may extract both using the 
sum rules for the splitting functions \cite{Gehrmann:2014yya}). We extracted these functions and found agreement
with the known results \cite{Stewart:2010qs, Ellis:1996nn, Furmanski:1980cm}, which serves as an additional check 
of our calculation.

\section{Numerics}
\label{sec:plots}

To illustrate the numerical impact of the NNLO corrections to the beam functions associated with the $I^\two_{ij}(z)$ computed here and in \mycite{Gaunt:2014xga}, we consider the integrated beam function
\begin{equation} \label{eq:tBFO}
 \tB_i(t_{\rm max},x,\mu_B) = \int^{t_{\rm max}}\! \df t\, B_i(t,x,\mu_B)
\,.\end{equation}
In all our numerical results, we pick a representative value of $t_{\rm max} = (30\GeV)^2$. The qualitative features in the numerical results only depend very little on the value of $t_{\rm max}$. We use the MSTW 2008 PDFs~\cite{Martin:2009iq} with their corresponding $\alpha_s(m_Z)$.

\begin{figure}[t]
\includegraphics[width=0.5\textwidth]{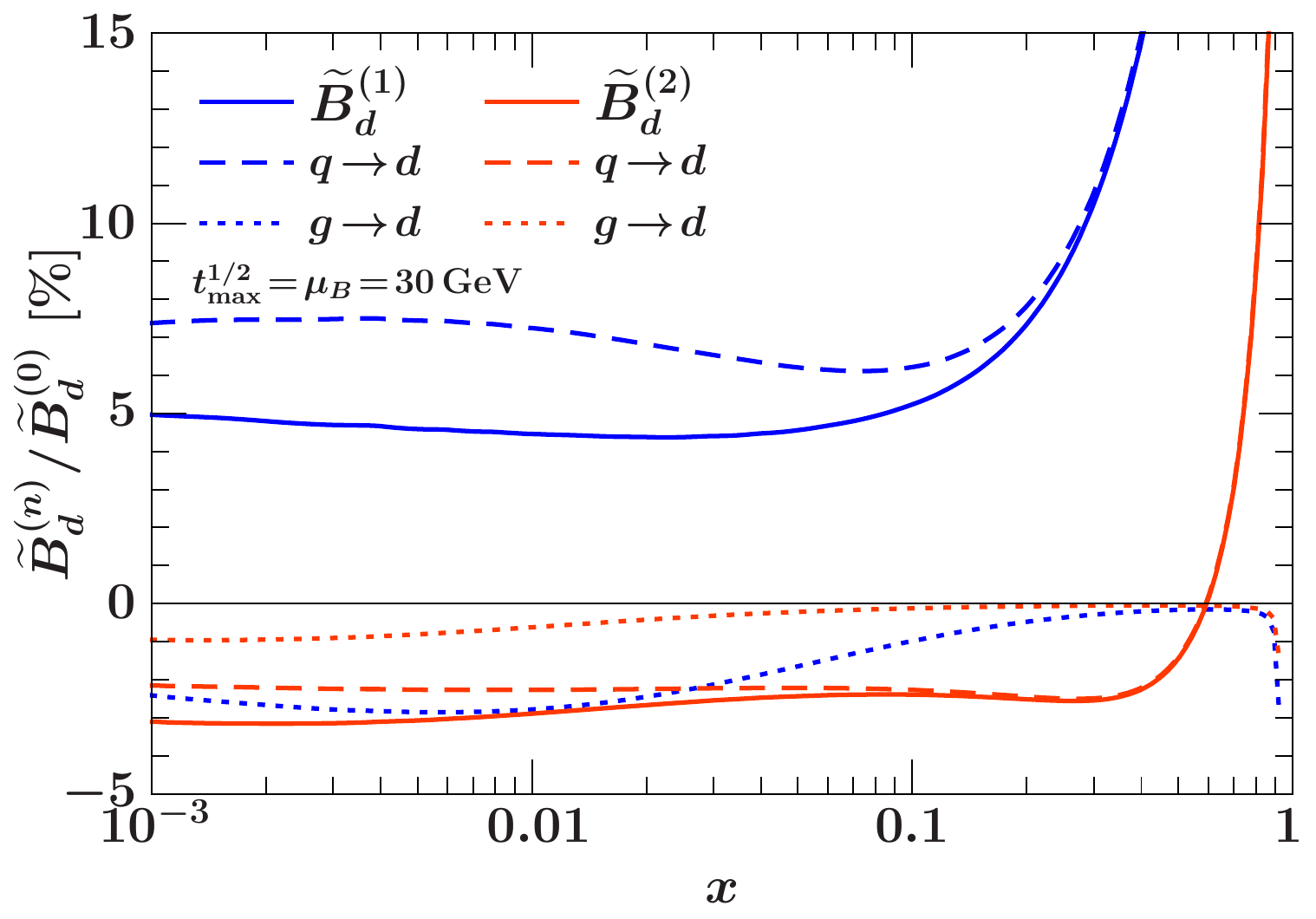}
\includegraphics[width=0.5\textwidth]{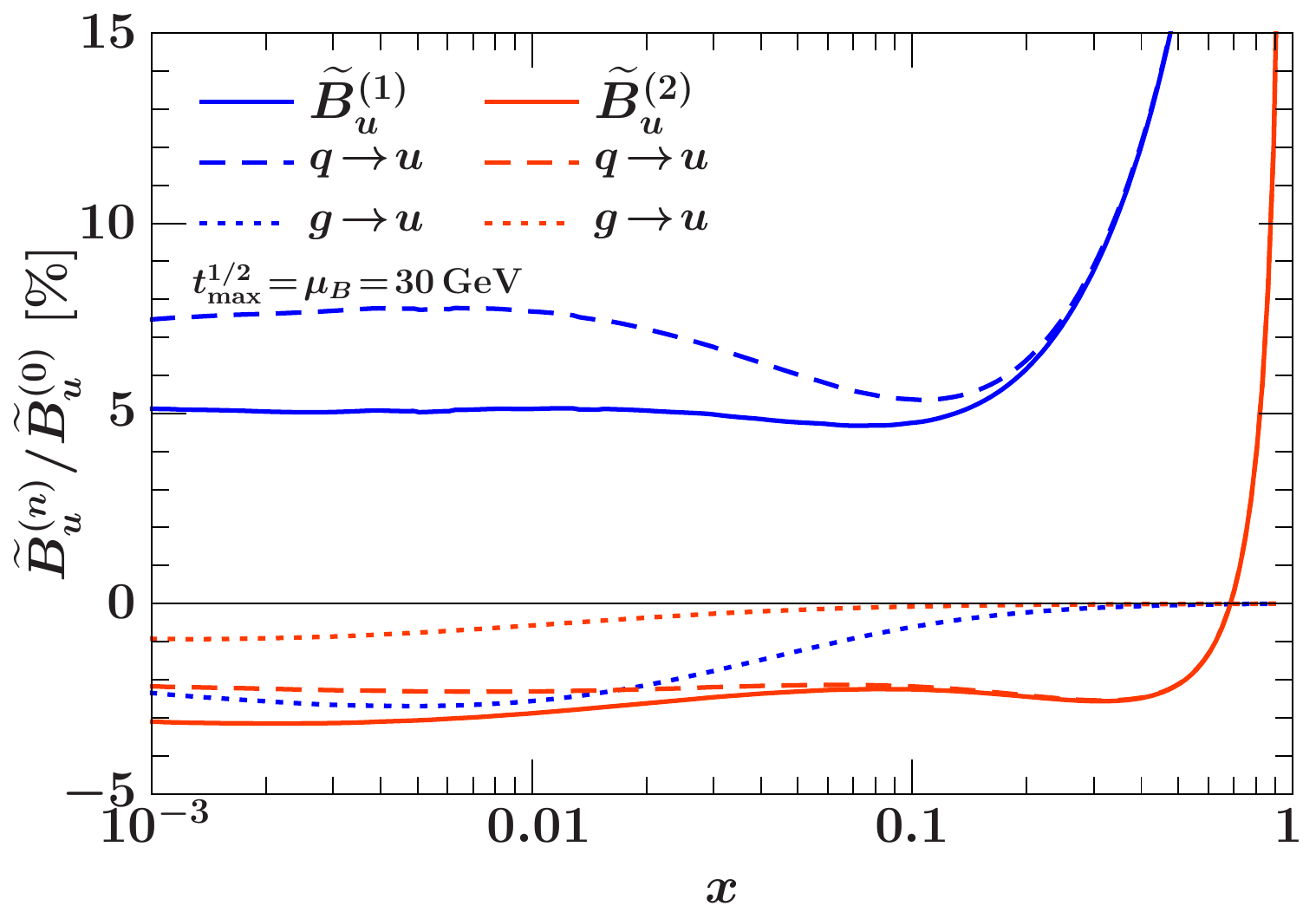}
\\
\includegraphics[width=0.5\textwidth]{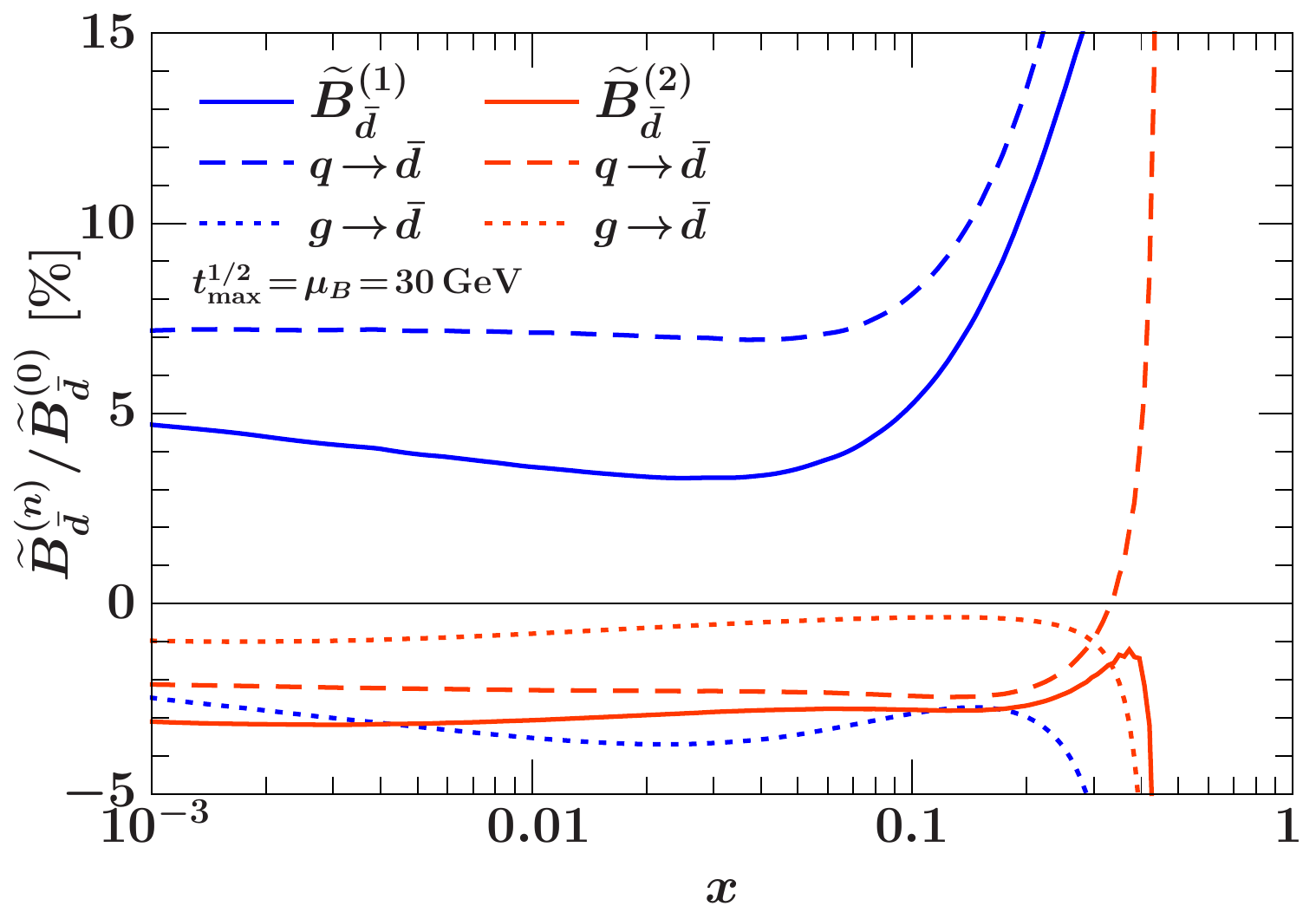}
\includegraphics[width=0.5\textwidth]{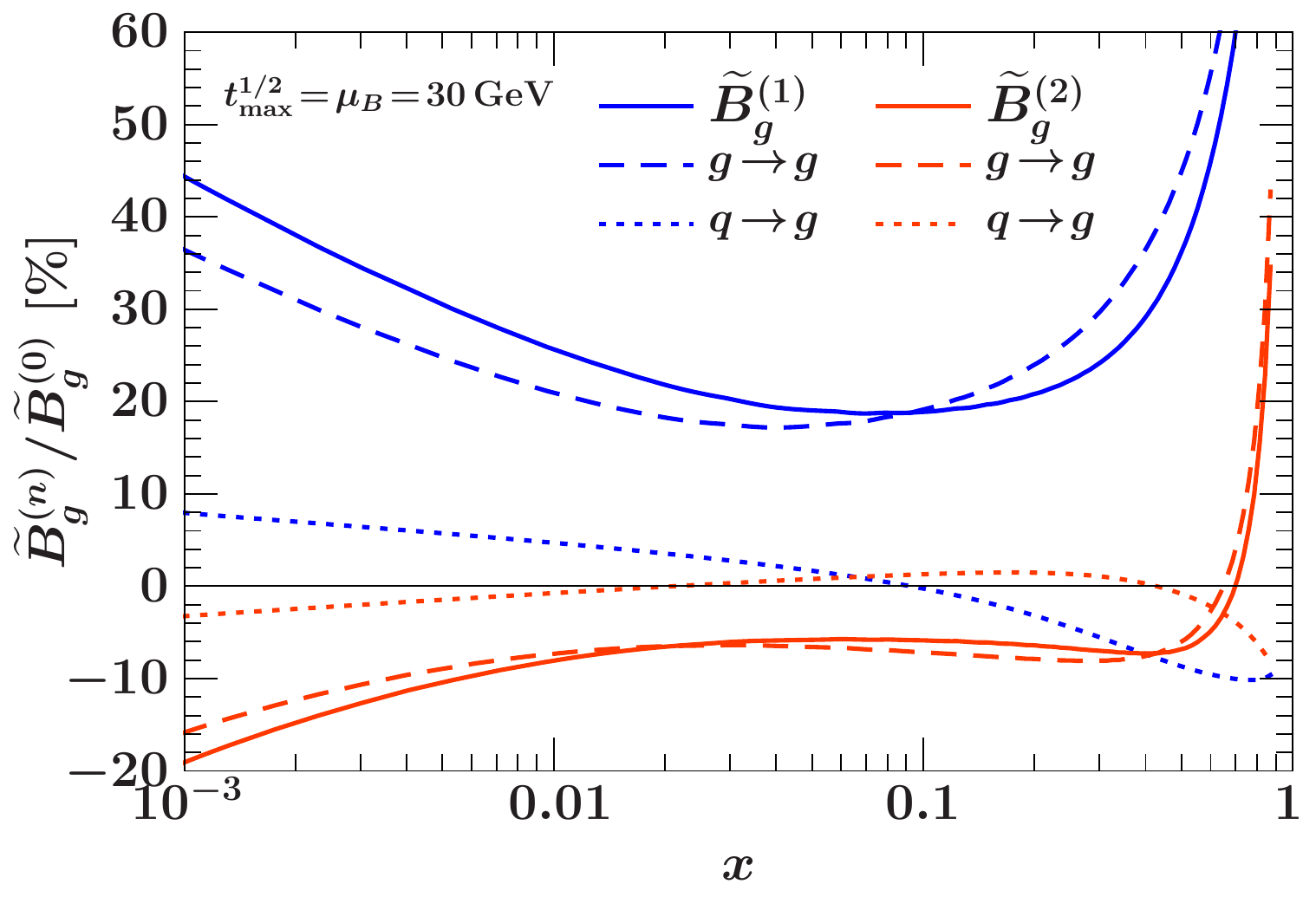}
\caption{
The one-loop (blue) and two-loop (orange) corrections to the integrated beam function $\tB_i$ in percent relative to the tree level result for $i=d$ (upper left), $i=u$ (upper right), $i=\bar d$ (lower left), and $i=g$ (lower right) as a function of the minus-momentum fraction $x$ carried by the parton $i$. Dotted lines show the contributions from off-diagonal channels ($q\to g$, $g \to q$), dashed lines the diagonal channels ($q\to q$, $g\to g$), as detailed in the text. The solid lines show the total result after summing over diagonal and off-diagonal contributions.
\label{fig:NNLOrel}}
\end{figure}

In \fig{NNLOrel}, we show the $\ord{\alpha_s}$ and $\ord{\alpha_s^2}$ contributions to $\tB_i(t_{\rm max},x,\mu_B)$ for $i=u,d,{\bar d},g$. For these plots we choose $\mu_B=\sqrt{t_{\rm max}} = 30\GeV$ such that all logarithms $\propto \ln^{n+1} (t_{\rm max}/\mu_B^2)$ from integrating the plus distributions $\cL_n(t/\mu_B^2)$ in \eq{I2master} vanish. Hence, with this scale choice, the $n$-loop correction to $\tB_i$ is directly given by
\begin{equation} \label{eq:tBFO2}
 \tB_i^{(n)}(t_{\rm max},x,\mu_B = \sqrt{t_{\rm max}})
 = \biggl[\!\frac{\alpha_s(\!\sqrt{t_{\rm max}})}{2\pi} \biggr]^{n} I^{(n)}_{ij}(x) \,\conv_x\, f_j(x,\sqrt{t_{\rm max}})\,,
\end{equation}
where $I^{(0)}_{ij}(z) = \delta_{ij}\delta(1-z)$ so at tree level $\tB_i^{(0)}(t_{\rm max},x,\mu_B = \sqrt{t_{\rm max}}) = f_i(x,\sqrt{t_{\rm max}})$.

For each parton $i=u,d,{\bar d},g$, \fig{NNLOrel} shows the pure one-loop correction, $\tB_i^{(1)}/\tB_i^{(0)}$ (blue) and the pure two-loop correction, $\tB_i^{(2)}/\tB_i^{(0)}$ (orange) in percent relative to the tree level result, as a function of the minus-momentum fraction $x$. Since here we care about the size of the terms in the perturbative series of the matching coefficients, we use the same NNLO PDFs everywhere.
For each order, we show three curves corresponding to the contributions from the diagonal (dashed lines), the off-diagonal (dotted lines), and the sum of all parton channels (solid lines). The diagonal contributions ($q\to d,u,\bar d$) to $\tB_{d,u, \bar d}$ include the sum of all possible (anti)quark-(anti)quark channels ($q_i, \bar q_i \to d,u,\bar d$). Similarly, the off-diagonal contribution ($q\to g$) to $\tB_g$ includes the sum over all (anti)quark-to-gluon contributions ($q_i, \bar q_i \to g$). Numerically, the corrections from the $q_i \to q_j$ channels in $\tB^\two_{q_j}$ with $i \neq j$ however turn out to be completely negligible.

In all cases in \fig{NNLOrel} we observe sizable negative total $\ord{\alpha_s^2}$ corrections compared to the positive total $\ord{\alpha_s}$ corrections. For the (anti)quark beam functions, the two-loop corrections are about half the size of the one-loop corrections. The reason is that the $g\to q$ mixing contribution is sizeable and always negative. As a result, at NLO it partially compensates the diagonal $q\to q$ contribution, reducing the absolute size of the total  $\ord{\alpha_s}$ correction. In contrast, at NNLO it adds to the diagonal contribution, enhancing the absolute size of the total $\ord{\alpha_s^2}$ correction. For the gluon beam function, the corrections are much larger than for the quark case, as expected from the larger color factor for gluons. Here, the off-diagonal mixing contributions $q\to g$ have a relatively smaller effect and add to the diagonal $g\to g$ contribution (for most of the relevant $x$ range). For $x\to 1$ the corrections become large due to the presence of threshold logarithms $\alpha_s^n \ln^{m}(1-x)$ in the diagonal terms. For $x \lesssim 0.2$, the corrections are largely independent of $x$, and their size complies with the typical pattern expected for perturbative QCD corrections at the considered scales.

\begin{table*}[t!]
  \centering
  \begin{tabular}{c | c c c | c c c}
  \hline \hline
  & \multicolumn{3}{c|}{matching} & \multicolumn{3}{c}{RGE running} \\
  order & $\cI_{ij}$ & PDF & $\alpha_s(m_Z)$ & $\gamma_{B}$ & $\Gamma_\mathrm{cusp}$ & $\beta$ \\ \hline
  NLL$'$ & NLO & NLO & $0.12018$ & $1$-loop & $2$-loop & $2$-loop \\
  NNLL & NLO & NLO & $0.12018$ & $2$-loop & $3$-loop & $3$-loop \\ \hline
  NNLL$'$ & NNLO & NNLO & $0.11707$ & $2$-loop & $3$-loop & $3$-loop \\
  N$^3$LL & NNLO & NNLO & $0.11707$ & $3$-loop & $4$-loop & $4$-loop \\
  \hline\hline
  \end{tabular}
  \caption{Perturbative ingredients entering at different orders in the resummed beam function.}
\label{tab:counting}
\end{table*}

\begin{figure}[t]
\includegraphics[width=0.5\textwidth]{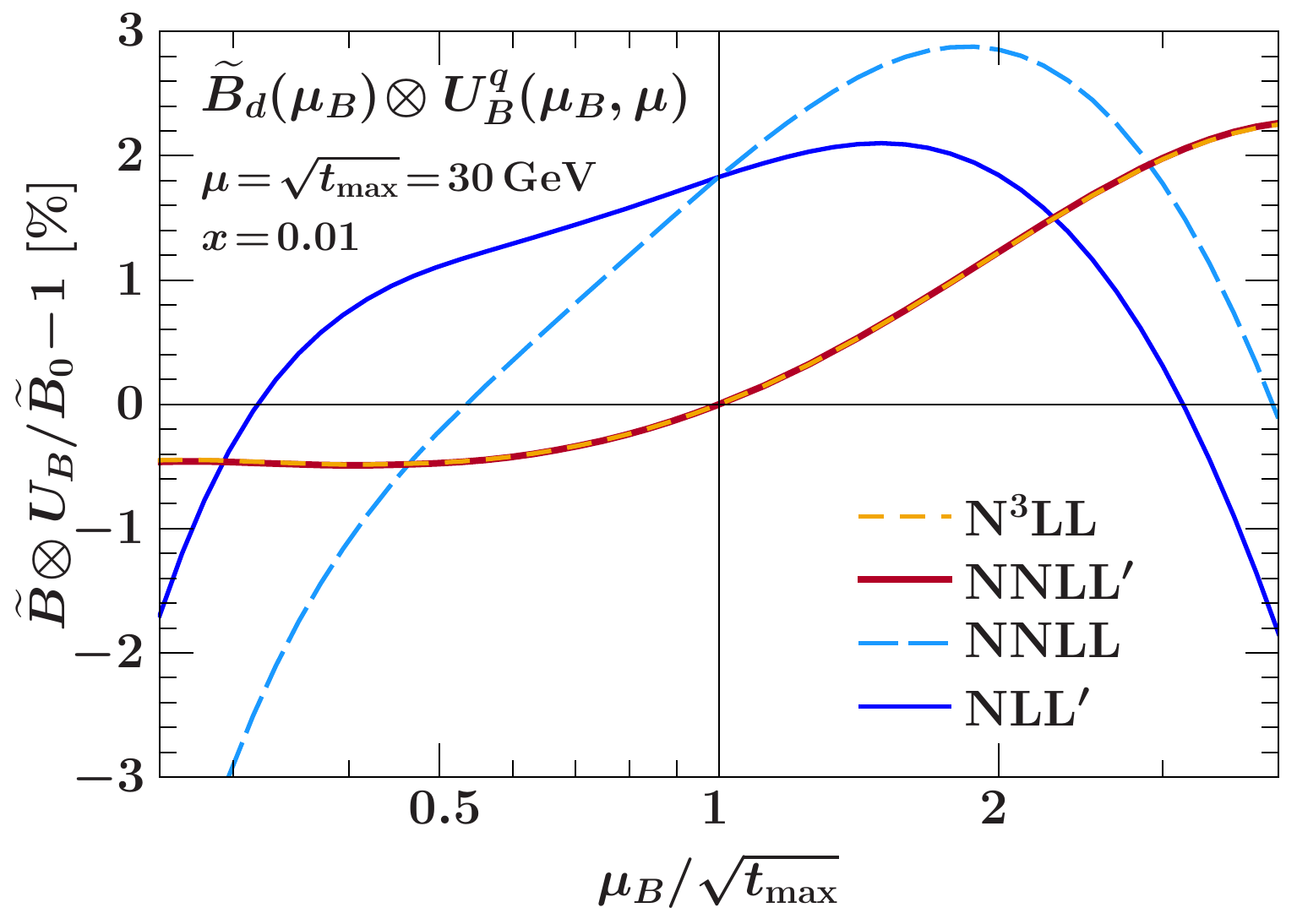}
\includegraphics[width=0.5\textwidth]{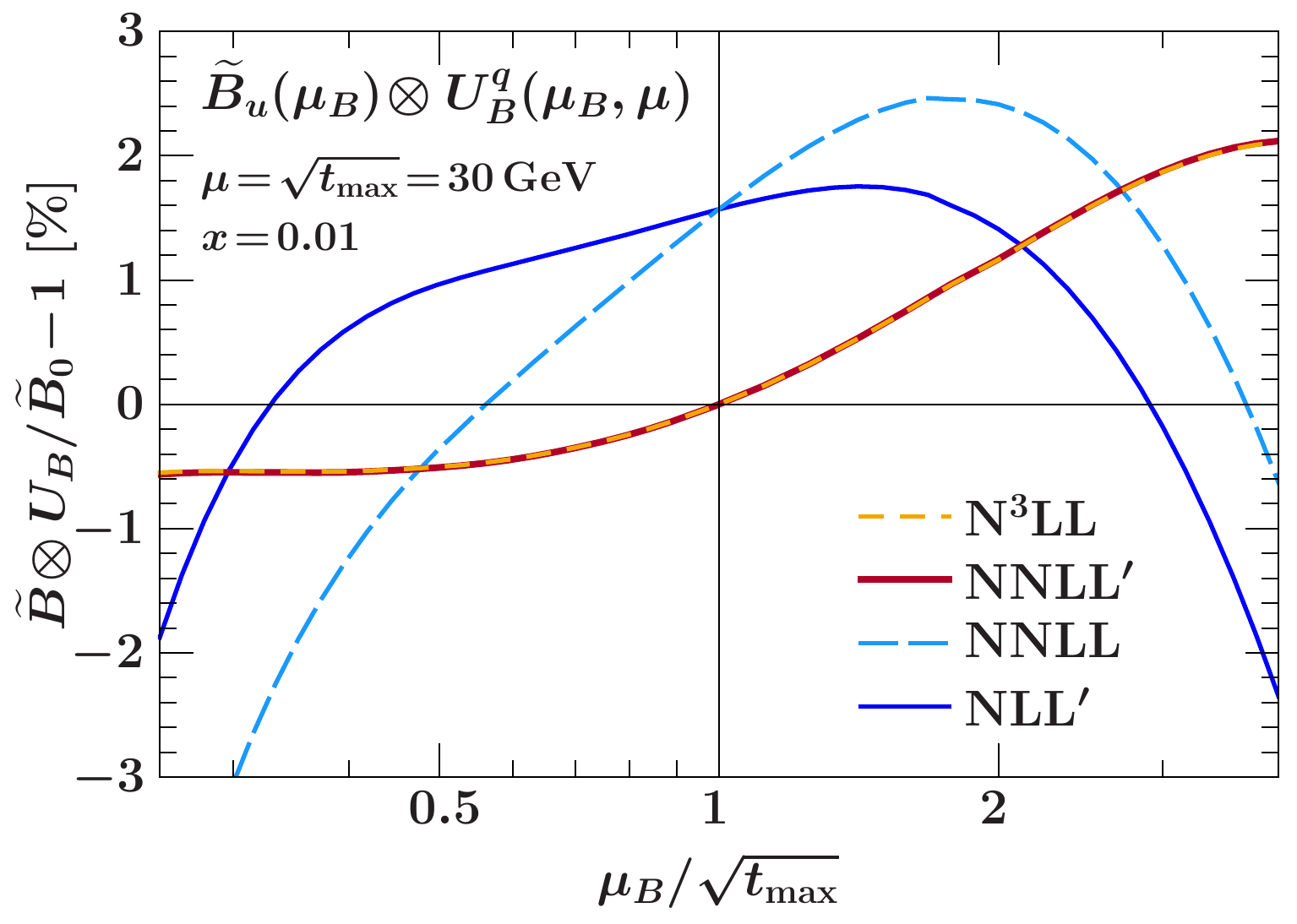}
\\
\includegraphics[width=0.5\textwidth]{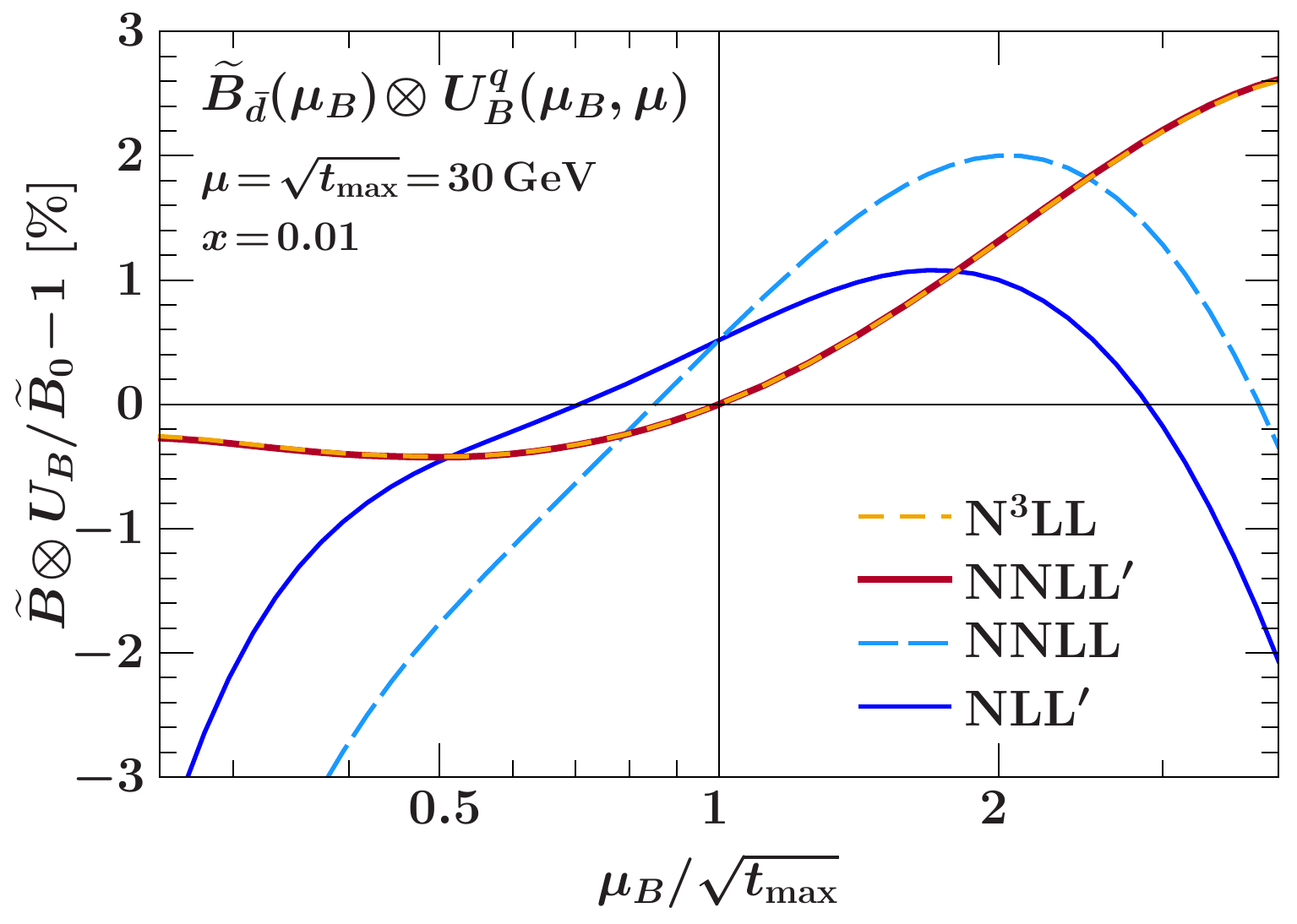}
\includegraphics[width=0.5\textwidth]{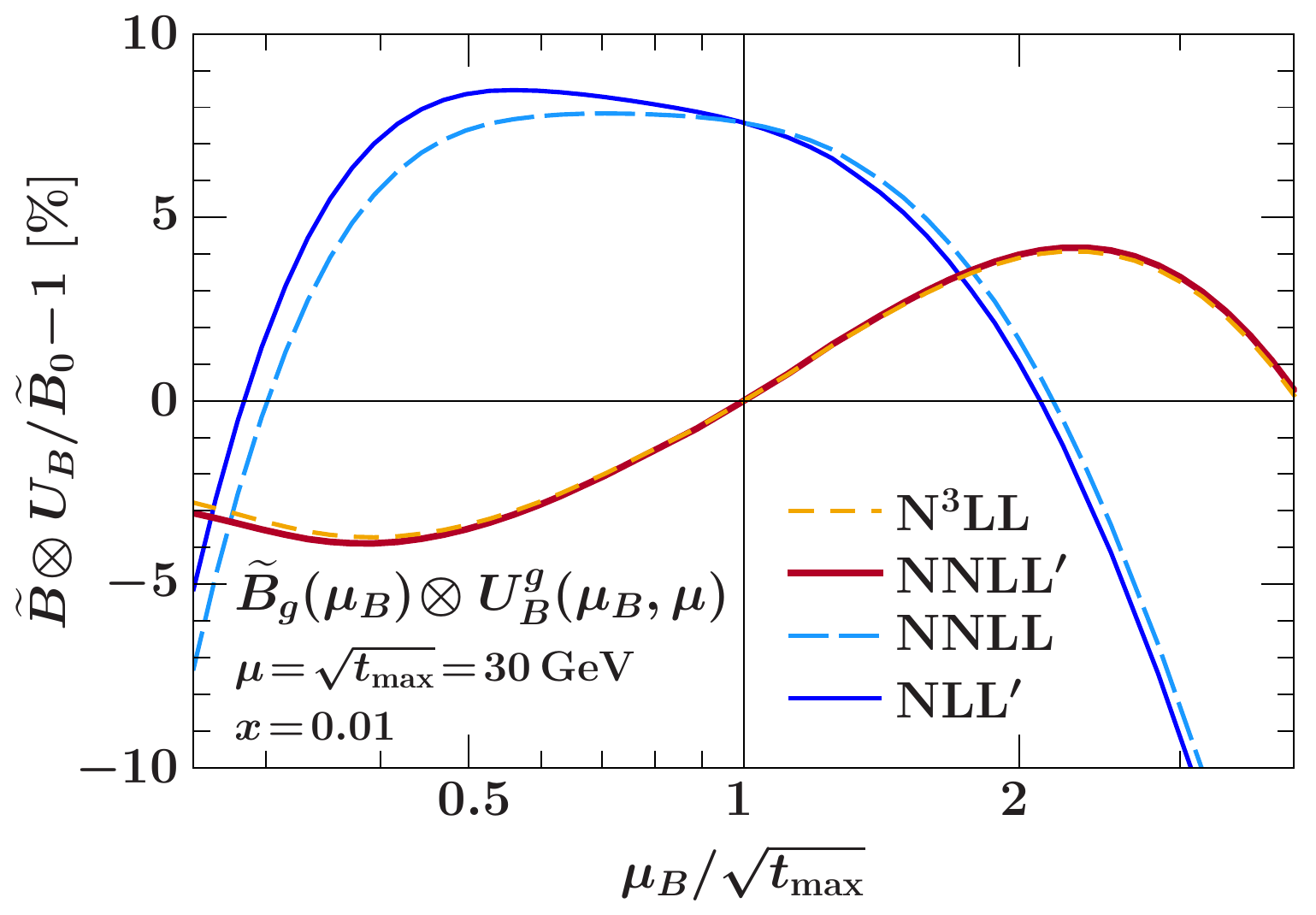}
\caption{
Residual matching scale dependence in the resummed integrated beam function $\widetilde B_i\otimes U_B$ at $x = 0.01$ for $i=d$ (upper left), $i=u$ (upper right), $i=\bar d$ (lower left), and $i=g$ (lower right). In all cases we show the correction in percent relative to the fixed NNLO result at the central scale $\widetilde B_{i0} \equiv \widetilde B_i^{\rm NNLO}(t_{\rm max}, x, \mu_B = \sqrt{t_{\rm max}})$.
\label{fig:resumscale}}
\end{figure}

To study the uncertainties in the perturbative series for the beam function, we cannot use a simple variation of $\mu_B$ in \eq{tBFO}, since the fixed-order beam function has an explicit dependence on $\mu_B$, containing Sudakov double logarithms in $\mu_B^2/t_{\rm max}$. Instead, we can consider the resummed beam function,
\begin{align} \label{eq:tBresum}
\tB_i(t_{\rm max},x,\mu)
&= \int^{t_{\rm max}}\! \df t\, \int\!\df t'\, B_i(t - t',x,\mu_B)\, U_B^i(t', \mu_B, \mu)
\nn \\
&= \int\! \df t'\, \widetilde B_i(t_{\rm max} - t',x,\mu_B)\, U_B^i(t', \mu_B, \mu)
\,,\end{align}
where the evolution factor $U_B^i$ only depends on $i = q$ or $i = g$ and can be found in \mycites{Stewart:2010qs, Berger:2010xi}.

The perturbative ingredients entering at a given resummation order are summarized in Table~\ref{tab:counting}. The one-loop matching enters at NLL$'$ and NNLL, while the two-loop matching enters at NNLL$'$ and N$^3$LL. The explicit expressions for the resummation kernels to N$^3$LL are taken from \mycite{Abbate:2010xh}. The noncusp anomalous dimensions are known~\cite{Stewart:2010qs, Berger:2010xi} from the three-loop results in \mycites{Moch:2004pa, Vogt:2004mw, Moch:2005id, Moch:2005tm}. The beta function is known up to four loops~\cite{Tarasov:1980au, Larin:1993tp, vanRitbergen:1997va}. The cusp anomalous dimension is known at present to three loops~\cite{Korchemsky:1987wg, Moch:2004pa}. For its four-loop coefficient formally needed at N$^3$LL, we use the Pade approximation $\Gamma_3 = \Gamma_2^2/\Gamma_1$. The numerical effect of varying $\Gamma_3$ by a factor of $\pm 3$ is much smaller than the effect induced by the known three-loop noncusp coefficient. To study the $\mu_B$ dependence in the resummed beam function, we use a consistent set of PDFs as required by the matching order, together with their corresponding $\alpha_s(m_Z)$ value, see Table~\ref{tab:counting}. The $\alpha_s$ running deserves some comment. Strictly speaking, the PDFs require three-loop (two-loop) running at NNLO (NLO), which is the same running order as required by the NNLL (NLL) RGE running. This means, the employed $\alpha_s$ running is formally fully consistent between PDFs and RGE at NNLL$'$ (NLL$'$) order, while at N$^3$LL (NNLL) order, the resummation requires the $\alpha_s$ running at four (three) loops, i.e., one order higher than the PDFs. In these cases, we use the higher $\alpha_s$ running, to have a fully consistent resummation, together with the numerical $\alpha_s(m_Z)$ value of the PDFs, which is the dominant effect as far as numerical consistency with the PDFs goes.%
\footnote{This is slightly different from the compromise used in \mycite{Berger:2010xi}, and seems to be the best possible compromise: Regarding the PDFs, the higher $\alpha_s$ running is formally a higher-order effect and numerically negligible, whereas the different $\alpha_s(m_Z)$ values required are by far a much larger numerical effect. On the other hand, it is formally needed to have a consistent RGE solution. While the numerical effect of the higher $\alpha_s$ running is very small, it is not negligible compared to other similarly small N$^3$LL running effects.}

The resummed $\tB_i(t_{\rm max},x,\mu)$ explicitly depends on the arbitrary scale $\mu$ but is formally independent of the matching scale $\mu_B$, with the $\mu_B$ dependence canceling between the fixed-order $\widetilde B_i(t_{\rm max},x,\mu_B)$ and the evolution factor $U_B(\mu_B, \mu)$ to the order one is working at. Hence, in \eq{tBresum}, we can use the residual dependence on the matching scale $\mu_B$ as an indication of the uncertainties due to missing higher-order corrections in the perturbative series of the beam function as long as $\mu_B^2 \simeq t_{\rm max}$ (so there are no large unresummed logarithms in the fixed-order series for $\widetilde B_i(t_{\rm max}, x, \mu_B)$). In \fig{resumscale} we show the residual matching scale dependence, varying $\mu_B$ between $\sqrt{t_{\rm max}}/4$ and $4\sqrt{t_{\rm max}}$, at different resummation orders, and for a representative value of $x = 0.01$. In these plots, we choose $\mu^2 = t_{\rm max}$ such that the central value at $\mu_B = \sqrt{t_{\rm max}}$ is equivalent to the pure NNLO or NLO result from \eq{tBFO2}. The purpose of the resummation here is thus not to resum large logarithms of $\mu/\mu_B$, but rather as a means to have a meaningful way to estimate the perturbative uncertainties from the residual matching scale dependence. All lines in these plots are shown as the percent change relative to the NNLO result at the central scale.

The NNLL (NLL) evolution factor cancels the explicit logarithmic $\mu_B$ dependence in the fixed-order $\widetilde B_i(t_{\rm max},x,\mu_B)$ to NNLO (NLO). Therefore, at NNLL$'$ (NLL$'$), shown by the solid lines in \fig{resumscale}, the $\mu_B$ dependence comes from both the residual $\mu_B$ dependence of the nonlogarithmic NNLO (NLO) matching corrections as well as the higher-order logarithmic corrections resummed by the evolution kernel. (The cancellation of the PDF scale dependence inside $\widetilde B_i(t_{\rm max},x,\mu_B)$ by both the diagonal and off-diagonal matching corrections also plays a nontrivial role.) By going to N$^3$LL (NNLL), shown by the dashed lines, one includes an additional uncanceled $\mu_B$ dependence from the three-loop (two-loop) noncusp anomalous dimension (as well as higher-order $\beta$ function and cusp pieces).

For the quark beam function, the effect of the two-loop noncusp is large and increases the overall $\mu_B$ variation at NNLL compared to NLL$'$. At N$^3$LL the opposite happens. Here, the three-loop noncusp corrections turn out to be tiny due to an accidental but almost perfect numerical cancellation in the combination $\gamma^q_{B\,2} - \gamma^q_{B\, 1} \beta_1/\beta_0$ that appears in the RGE solution. The effect of the four-loop cusp is tiny, which is not unusual. This overall pattern is consistent with that seen in \mycite{Abbate:2010xh} for thrust to N$^3$LL, which has an equivalent resummation structure. For the gluon beam function, the same cancellation does not happen, so the effect of the three-loop noncusp at N$^3$LL is visible and consistent with the size of the NNLL effect suppressed by an additional power of $\alpha_s$. (The numerical effect of the four-loop cusp is tiny here as well.) Overall, the scale dependence is larger in the gluon case due to the larger color factor for gluons than quarks.

Including the two-loop matching corrections reduces the matching scale dependence by a factor of two, from $2-3\%$ to $1-1.5\%$ for quarks and $\sim 8\%$ to $\sim 4\%$ for gluons. In complete resummed cross sections, the perturbative uncertainties due to the beam function component are often evaluated by separately varying the beam function scale, see e.g. \mycites{Stewart:2010pd, Berger:2010xi, Jouttenus:2013hs, Kang:2013wca, Liu:2013hba, Kang:2013nha, Kang:2013lga, Stewart:2013faa, Li:2014ria}. This is typically implemented through profile scale variations~\cite{Ligeti:2008ac, Abbate:2010xh}, which in the resummation region correspond to the canonical beam scales we have used here. In this context, this can be an important source of uncertainties. For example, in \mycite{Berger:2010xi} the gluon beam function gives the largest uncertainty in the resummation regime. We emphasize that final conclusions concerning the perturbative convergence and uncertainties can of course only be drawn by looking at the cross section for physical observables. Nevertheless, the overall reduction in the matching scale dependence in the resummed beam function at NNLL$'$ and N$^3$LL gives a good indication of the possible reduction in uncertainties in resummed predictions.

\section{Conclusions} 
\label{sec:conclusions}

In this paper, we have completed the calculation of the NNLO virtuality-dependent beam functions $B_{i}(t, x, \mu)$, by computing the gluon matching coefficients $\cI_{gj}(t,z,\mu)$ for the gluon beam function onto the PDFs at two loops. These results are an important ingredient to obtain the full NNLO singular contributions as well as the NNLL$'$ and N$^3$LL resummation for observables that probe the virtuality of the colliding partons, such as beam thrust and $N$-jettiness.

The methodology used here is the same as in our previous calculation of the two-loop quark matching
coefficients in \mycite{Gaunt:2014xga}. As in the quark case, we have checked our calculation by using two different
gauges -- Feynman and axial light-cone gauge -- and two different methods for taking the discontinuities 
of the operator diagrams that are required to obtain the partonic beam function matrix elements.
Our calculation provides an explicit verification at two loops of the all-orders result~\cite{Stewart:2010qs} that the
beam and jet function anomalous dimensions are equal also for the gluon case. Conversely, relying on this fact, we are able 
to extract the two-loop gluon splitting functions, $P_{gi}$, and find agreement with the well-known results~\cite{Ellis:1996nn, Furmanski:1980cm}.

We have also presented numerical results for the (anti)quark and gluon beam functions at NNLO as well as for the resummed beam function to N$^3$LL. We find that the numerical effects of the two-loop corrections are important. They are about half the size of the one-loop corrections but with opposite sign (except near $x= 1$). For the resummed beam function, the residual dependence on the matching scale $\mu_B$ gives an indication of the perturbative uncertainties due to missing higher order corrections. It reduces by roughly a factor of two when our new two-loop matching corrections are included.

\begin{acknowledgments}
The Feynman diagrams in this paper have been drawn using {\tt JaxoDraw}~\cite{Binosi:2008ig}.
Parts of the calculations in this paper and \cite{Gaunt:2014xga} were perfomed using {\tt FORM}
 \cite{DBLP:journals/corr/abs-1203-6543}, {\tt HypExp} \cite{Huber:2005yg, Huber:2007dx} and {\tt FeynCalc} \cite{Mertig:1990an}.
This work was supported by the DFG Emmy-Noether Grant No. TA 867/1-1.
\end{acknowledgments}

\appendix

\section{Perturbative ingredients}
\label{app:pert}

In this appendix we summarize the additional perturbative ingredients for the gluon beam function. These have been given previously in \mycite{Berger:2010xi} and are repeated here for completeness.

The coefficients of the cusp, noncusp, and PDF anomalous dimensions are defined according to
\begin{align}
\gamma^i_B(\alpha_s) &= \sum_{n=0}^\infty  \Bigl(\frac{\alpha_s}{4\pi}\Bigr)^{n+1} \gamma_{B\,n}^i
\,, \qquad
\Gamma^i_\cusp(\alpha_s) = \sum_{n=0}^\infty  \Bigl(\frac{\alpha_s}{4\pi}\Bigr)^{n+1} \Gamma^i_n
\,, \nn \\
& \qquad P_{ij}(z,\alpha_s) = \sum_{n=0}^{\infty} \left(\dfrac{\alpha_s}{2\pi}\right)^{n+1} P_{ij}^{(n)}(z)
\,.\end{align}
The $\overline{\mathrm{MS}}$ anomalous dimension coefficients for the gluon beam function up to three loops are
\begin{align} \label{eq:gammaBgexp}
\gamma_{B\,0}^g &= 2 \beta_0
\,,\nn\\
\gamma_{B\,1}^g
&= C_A \Bigr[
   C_A \Bigl(\frac{182}{9} - 32\zeta_3\Bigr)
   + \beta_0 \Bigl(\frac{94}{9}-\frac{2\pi^2}{3}\Bigr) \Bigr]
   + 2\beta_1
\,,\nn\\
\gamma_{B\,2}^g
&= C_A \Bigr[
   C_A^2 \Bigl(\frac{49373}{81} - \frac{944 \pi^2}{81} - \frac{16\pi^4}{5} - \frac{4520 \zeta_3}{9}
      + \frac{128\pi^2 \zeta_3}{9} + 224 \zeta_5 \Bigr)
   \nn \\ & \qquad
   +  C_A \beta_0 \Bigl(-\frac{6173}{27} - \frac{376 \pi^2}{81} + \frac{13\pi^4}{5} + \frac{280\zeta_3}{9}\Bigr)
   +  \beta_0^2 \Bigl(-\frac{986}{81}-\frac{10\pi^2}{9}+\frac{56 \zeta_3}{3}\Bigr)
   \nn \\ & \qquad
   +  \beta_1 \Bigl(\frac{1765}{27} - \frac{2\pi^2}{3} - \frac{8\pi^4}{45} - \frac{304\zeta_3}{9}\Bigr) \Bigr]
   + 2 \beta_2
\,. \end{align}
The coefficients of the cusp anomalous dimension and beta function are given in \mycite{Gaunt:2014xga}.

The one-loop gluon matching coefficients appearing in \eq{I1master} are written as
\begin{align}
I_{gg}^\one(z) &= C_A\, \theta(z) I_{gg}(z)
\,,\nn\\
I_{gq_i}^\one(z) &= C_F\, \theta(z) I_{gq}(z)
\,,\end{align}
with the one-loop matching functions\footnote{Note that $I_{ij}(z) \equiv \cI_{ij}^{(1,\delta)}(z)$ in the notation of \mycites{Stewart:2010qs, Berger:2010xi}.}
\begin{align} \label{eq:Igdel_results}
I_{gg}(z)
&= \cL_1(1-z)\,\frac{2(1-z + z^2)^2}{z} - \frac{\pi^2}{6} \delta(1-z) - P_{gg}(z) \ln z
\,, \nn \\
I_{gq}(z)
&= P_{gq}(z)\ln \frac{1-z}{z} + \theta(1-z) z
\,.\end{align}

The one-loop PDF anomalous dimension in the $\overline{\mathrm{MS}}$ scheme are
\begin{align}
P_{gg}^\zero(z) &= C_A\, \theta(z) P_{gg}(z) + \frac{\beta_0}{2}\,\delta(1-z)
\,,\nn\\
P_{gq_i}^\zero(z) = P_{g\bar q_i}^\zero(z) &= C_F\, \theta(z) P_{gq}(z)
\,,\end{align}
with the LO gluon splitting functions
\begin{align} \label{eq:Pij}
P_{gg}(z)
&= 2 \cL_0(1-z) \frac{(1 - z + z^2)^2}{z}
\,,\nn\\
P_{gq}(z) &= \theta(1-z)\, \frac{1+(1-z)^2}{z}
\,.\end{align}
At two loops we write
\begin{align}
P_{gg}^\one(z) &= \theta(z) \bigl[ C_A P_{ggA}^\one(z) + T_F n_f P_{ggF}^\one(z) \bigr]
\,, \nn \\
P_{g q_i}^\one(z) = P_{g \bar q_i}^\one(z) &= C_F\, \theta(z) P_{gq}^\one(z)
\,,\end{align}
where the NLO gluon splitting functions are given by~\cite{Furmanski:1980cm, Ellis:1996nn}
\begin{align} \label{eq:Pgg_one}
P_{ggA}^\one(z)
&=  \frac{\Gamma_1}{8} P_{gg}(z) + \delta(1 - z)\bigl[ C_A (-1 + 3 \zeta_3) + \beta_0 \bigr]
   \nn \\ & \quad
   + C_A \biggl\{
      P_{gg}(z) \Bigl[-2\ln(1-z) + \frac{1}{2} \ln z \Bigr] \ln z
      + P_{gg}(-z) \Bigl[S_2(z) + \frac{1}{2}\ln^2 z \Bigr]
      \nn\\ & \qquad
      + 4(1 + z) \ln^2 z - \frac{4(9 + 11 z^2)}{3}  \ln z
      - \frac{277}{18 z} + 19(1 - z) + \frac{277}{18} z^2
   \biggr\}
   \nn\\ & \quad
   + \beta_0 \Bigl[\frac{13}{6z} - \frac{3}{2}(1 - z) - \frac{13}{6} z^2 + (1 + z) \ln z \Bigr]
\,,\nn\\
P_{ggF}^\one(z)
&= C_F \Bigl[-\delta(1-z) + \frac{4}{3 z} - 16 + 8 z + \frac{20}{3} z^2 - 2(1+z) \ln^2 z -2 (3 + 5 z) \ln z \Bigr]
\,,\end{align}
and
\begin{align}
P_{gq}^\one(z)
&= C_A \biggl\{
   P_{gq}(z) \Bigl[\ln^2(1-z) - 2\ln (1-z) \ln z  - \frac{101}{18} - \frac{\pi^2}{6} \Bigr]
   + P_{gq}(-z) S_2(z)
   \nn\\ & \qquad
   + 2 z \ln(1-z)
   + (2+z) \ln^2z
   - \frac{36 + 15z + 8 z^2}{3} \ln z
   + \frac{56 - z + 88 z^2}{18}
   \biggr\}
   \nn\\ & \quad
   - C_F \biggl\{
      P_{gq}(z) \ln^2(1-z) + [3 P_{gq}(z) + 2z]\ln(1-z)
      + \frac{2 - z}{2} \ln^2 z - \frac{4 + 7 z}{2} \ln z
      \nn\\ & \qquad
      + \frac{5 + 7 z}{2}
   \biggr\}
   + \beta_0 \Bigl\{ P_{gq}(z) \Bigl[\ln(1-z) + \frac{5}{3}\Bigr] + z \Bigr\}
\,.\end{align}

The Mellin convolution of two functions is defined as (where the index $j$ is not summed)
\begin{equation}
(P_{ij} \conv P_{jk})(z) \equiv P_{ij}(z) \convz P_{jk}(z) = \int_z^1\! \frac{\df w}{w}\, P_{ij}(w) P_{jk} \Bigl(\frac{z}{w}\Bigr)
\,.\end{equation}
The convolutions of two one-loop QCD splitting functions for a final gluon are
\begin{align}
(P_{gg} \conv P_{gg})(z)
&= 8 \cL_1(1-z)\frac{(1 - z + z^2)^2}{z} - \frac{2\pi^2}{3} \delta(1-z)
  - 2[P_{gg}(z) + 4(1+z)]\ln z
  \nn \\ & \quad
  - \frac{44}{3z} + 12(1-z) + \frac{44}{3}z^2
\,, \nn \\
(P_{gq} \conv P_{qg})(z)
&= 2(1+z)\ln z
  + \frac{4}{3z} + 1 - z - \frac{4}{3}z^2
\,, \nn \\
(P_{gq} \conv P_{qq})(z)
&= 2 P_{gq}(z) \ln(1-z)
  + (2-z) \ln z
  + 2 - \frac{z}{2}
\,, \nn \\
(P_{gg} \conv P_{gq})(z)
&= 2 P_{gq}(z) \ln\frac{1-z}{z}
  - 2(4 + z) \ln z
  - \frac{31}{3z} + 8 + z + \frac{4}{3} z^2
\,.\end{align}
The convolutions of the one-loop gluon matching functions with the one-loop splitting functions are
\begin{align} \label{eq:convres}
(I_{gg} \conv P_{gg})(z)
&= 6\cL_2(1-z)\frac{(1-z+z^2)^2}{z} + 4 \zeta_3 \delta(1-z)
  + P_{gg}(z) \Bigl[\ln^2 z - 4\ln(1-z)\ln z - \frac{\pi^2}{2}\Bigr]
  \nn \\ & \quad
  + 8 (1 + z) \Bigl[\Li_2(z) + \frac{1}{2} \ln^2 z - \frac{\pi^2}{6} \Bigr]
  + \Bigl(-\frac{22}{3z} + 14 - 4 z + \frac{44}{3}z^2\Bigr) \ln\frac{1-z}{z}
  \nn \\ & \quad
  - \Bigl(\frac{22}{3z} + 2 + 8z\Bigr) \ln(1-z)
  + \frac{67}{9z} -  \frac{23}{3}(1-z) - \frac{67}{9}z^2
\,, \nn \\
(I_{gq} \conv P_{qg})(z)
&= - 2(1 + z)\Bigl[\Li_2(z) + \frac{1}{2} \ln^2 z - \frac{\pi^2}{6} \Bigr]
  + \Bigl(\frac{4}{3z} - 3z - \frac{4}{3}z^2\Bigr)\ln\frac{1-z}{z}
  \nn \\ & \quad
  + (1 + 2 z)\ln(1-z)
  - \frac{13}{9z} + \frac{4}{3} + \frac{2}{3}z-\frac{5}{9}z^2
\,, \nn \\
(I_{gq} \conv P_{qq})(z)
&= 2 P_{gq}(z) \Bigr[\ln(1-z) \ln\frac{1-z}{z}-\frac{\pi^2}{6} + \frac{5}{8}\Bigr]
  - (2 - z) \Bigl[\Li_2(z) + \frac{1}{2} \ln^2 z - \frac{\pi^2}{6} - \frac{1}{4} \Bigr]
  \nn \\ & \quad
  + \frac{4+3z}{2} \ln (1-z) - \frac{2+z}{2} \ln z
 \,, \nn \\
(I_{gg} \conv P_{gq})(z)
&= P_{gq}(z) \Bigl(\ln^2\frac{1-z}{z} - \frac{\pi^2}{6}\Bigr)
   + 2(4+z) \Bigl[\Li_2(z) + \frac{1}{2} \ln^2 z - \frac{\pi^2}{6} \Bigr]
  + \frac{21 - 26 z + 5 z^2}{6z}
  \nn \\ & \quad
  + \Bigl(-\frac{3}{z} + 10 + 3z + \frac{4}{3}z^2\Bigr) \ln\frac{1-z}{z}
  - \Bigl(\frac{22}{3z} + 2 + 2z \Bigr)\ln(1-z)
\,.\end{align}

\section{Change of transverse variables in the On-Shell Diagram method}
\label{sec:TChange}

In this appendix we describe and motivate the change of transverse variables employed
to calculate the `real-real' cuts of the diagrams in the On-Shell Diagram Method.
This method involves taking the discontinuities of the
diagrams at the very beginning by taking all possible cuts of the diagrams using the Cutkosky
rules~\cite{Cutkosky:1960sp, Veltman:1994Ve}, and by
`real-real' cuts we mean cuts that do not leave any virtual loops on either side of the cut.

\begin{figure}
\centering
\includegraphics[width=0.3\textwidth]{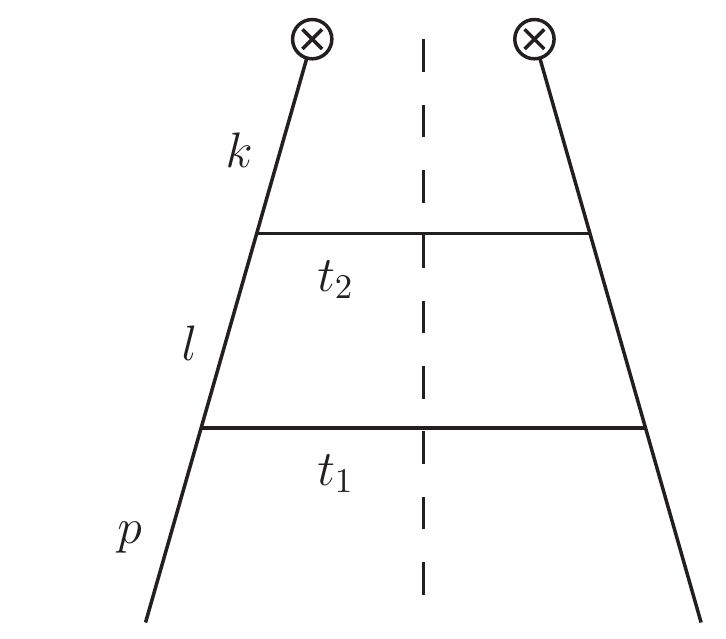}
\caption{The cut ladder diagram. The cut is denoted by a dashed line. The discussion
in this section applies regardless of the species of partons in the diagram, so we just
use straight lines to denote the particles in the diagram.}
\label{fig:TMchangeex}
\end{figure}

Let us take as an example `real-real' diagram the ladder diagram of \fig{TMchangeex}, where we have
drawn the cut on the diagram and indicated the momentum for each line. We decompose the on-shell momenta $t_1$ and $t_2$ as follows:
\begin{equation} \label{eq:LCdecomp}
t_i = \dfrac{z_i p^-}{2}\, n + \dfrac{\mathbf{t}_i^2}{2z_ip^-}\, \bar{n} + t_{iT}
\,,\end{equation}
where $n$ is a dimensionless light-cone vector pointing along $p$, and $\bar{n}$ is another 
dimensionless light-cone vector satisfying  $n\cdot \bn=2$.
$t_{iT}$ is a transverse vector satisfying $t_{iT} \cdot \bn=t_{iT} \cdot n=0$, 
$t_{iT} \cdot t_{jT} \equiv -\mathbf{t}_i \cdot \mathbf{t}_j$.

One nontrivial integral we have to evaluate for \fig{TMchangeex} has the structure
\begin{equation} \label{eq:CutIntegral1}
\int\!\df\Phi(t_1, t_2)\, \delta\Bigl(z-\dfrac{\bar{n}\cdot k}{\bar{n} \cdot p} \Bigr) \delta\Bigl(\frac{t}{2z} + p\cdot k \Bigr) \frac{f(z_1,z_2)}{l^2\, k^2}
\,,\end{equation}
where $\df\Phi(t_1, t_2)$ is the on-shell phase space for the cut particles,
and $f(z_1,z_2)$ is some function that we will not concern ourselves
with here.

Writing \eq{CutIntegral1} in terms of the on-shell momenta, and decomposing these
according to \eq{LCdecomp}, we obtain the following result for the transverse part
of the integral (this expression is then multiplied by some function of $z_1$ and $z_2$
and finally integrated over $z_1$ and $z_2$):
\begin{equation} \label{eq:CutIntegral2}
\int\! \df^{d-2}\mathbf{t}_1\, \df^{d-2}\mathbf{t}_2\,
\delta\biggl(\frac{t}{2z} - \frac{\mathbf{t}_1^2}{2z_1} - \frac{\mathbf{t}_2^2}{2z_2} \biggr)\, \frac{-z_1}{\mathbf{t}_1^2}\;
\frac{-1}{t + (\mathbf{t}_1+\mathbf{t}_2)^2}
\,.\end{equation}

We would like to make a change of transverse variables that simplifies the denominators
and delta function argument in \eq{CutIntegral2}, such that the integral is easier
to perform. Ideally, one would like to remove the dependence on the angle from these quantities
such that the angular integral is trivial, but in the case of \eq{CutIntegral2}
that is not possible. Instead, we perform a change of variables such that in terms of the 
transformed variables $\mathbf{r_1}$ and $\mathbf{r}_2$, one denominator is of the form 
$(\mathbf{r}_1 + \mathbf{r}_2)^2$, whilst the other denominator and the argument of the 
delta function do not depend on the angle between $\mathbf{r}_1$ and $\mathbf{r}_2$. The choice of variables we use is
\begin{align} \label{eq:Tvarchange}
\mathbf{t}_1 = \mathbf{r}_1 + \mathbf{r}_2\,, \qquad  \mathbf{t}_2 = -\mathbf{r}_1 + \frac{z_2}{z_1}\mathbf{r}_2
\,.\end{align}
Then the integral \eqref{eq:CutIntegral2} becomes
\begin{align} \label{eq:CutIntegral3}
&(-z_1)\left(\dfrac{1-z}{z_1}\right)^{2-2\epsilon}\dfrac{\pi^{1-\epsilon}}{\Gamma(1-\epsilon)} \dfrac{\pi^{\tfrac{1}{2}-\epsilon}}{\Gamma(\tfrac{1}{2}-\epsilon)}
\int\! r_1^{-2\epsilon} r_2^{-2\epsilon} \df r^2_1\, \df r^2_2\, \int_0^\pi\!\df\theta\, \sin^{-2\epsilon}(\theta)
\nn \\
&\quad\;\times
\delta\biggl[\frac{t}{2z}-\dfrac{1-z}{2z_1z_2}\Bigl(r_1^2+r_2^2\dfrac{z_2}{z_1}\Bigr)\biggr]
\frac{1}{t+(1-z)^2r_2^2/z_1^2}\, \frac{1}{(\mathbf{r}_1 + \mathbf{r}_2)^2}
\,.\end{align}
where $d=4-2\epsilon$, $r_i = |\mathbf{r}_i|$, and $\theta$ is the angle between $\mathbf{r}_1$ and $\mathbf{r}_2$.
This change of variables allows us to use
\begin{equation}
\int_0^\pi\!\df\theta\, \frac{\sin^{-2\epsilon}(\theta)}{(\mathbf{a}+\mathbf{b})^2}
= \dfrac{1}{\mathbf{b}^2}
B\Bigl(\frac{1}{2}-\epsilon, \frac{1}{2}\Bigr) \,_2F_1\Bigl( 1,1+\epsilon;1-\epsilon; \frac{\mathbf{a}^2}{\mathbf{b}^2}\Bigr)
\qquad \text{for}\qquad
\mathbf{a}^2 < \mathbf{b}^2
\end{equation}
to evaluate the integral over $\theta$ \cite{gradshteyn2007},
where $B(x,y)$ is the beta function, and ${}_2F_1(a,b;c;x)$ the Gaussian hypergeometric
function. Similar reasoning is behind the change of transverse variables performed in Appendix B of 
\mycite{Ellis:1996nn}, though in that paper there is a delta function fixing $k^2$ rather than
$n\cdot k$ as we have here.

After the angular integral, we have two terms, one of which corresponds to $r_1 < r_2$ and 
the other of which corresponds to $r_2 < r_1$. The integrals over one of the magnitudes $r_1$
or $r_2$ in these terms can be performed using the delta function of $t$, whilst the other can
be performed using a straightforward variable transform (often only as an expansion in $\epsilon$). 
Then all that remains are the integrals over the components $z_1$ and $z_2$, which can be performed using standard techniques.

The change of transverse variables in \eq{Tvarchange} is sufficient to evaluate all
other nontrivial integrals for the ladder diagram, and indeed the integrals for all 
other topologies in the two-loop calculation.

\section{Virtual integral containing a light-cone divergence}
\label{sec:LCint}

\begin{figure}
\centering
\includegraphics[width=0.7\textwidth]{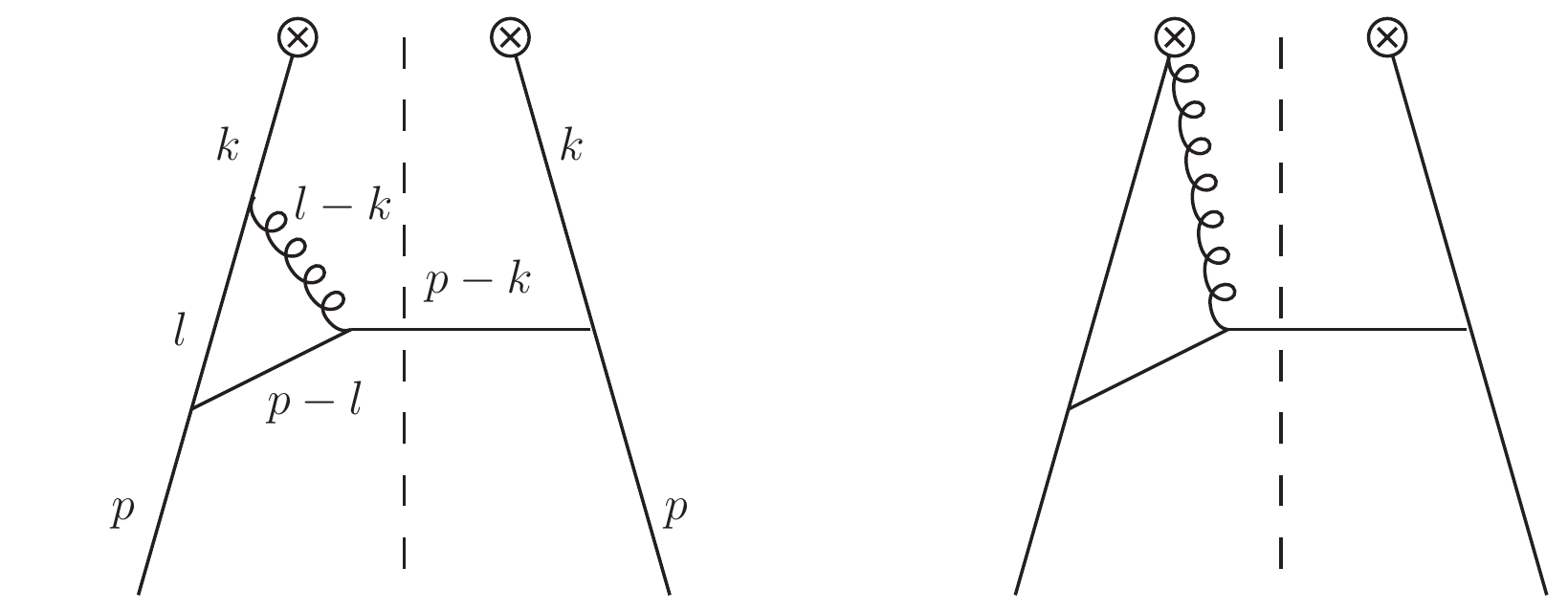}
\put(-300,105){(a)}
\put(-125,105){(b)}
\caption{Cut graph that contains the virtual three-point integral \ref{eq:LCint1} in (a) light-cone
gauge and (b) Feynman gauge. The straight lines can be any type of parton.}
\label{fig:threepoint}
\end{figure}

The integral in which we cannot simply use dimensional regularization
to regulate all divergences is the virtual three-point integral
\begin{align} \label{eq:LCint1}
\int\! \dfrac{\df^dl}{(2\pi)^d}\, \dfrac{\bn\!\cdot\! p}{l^2\, (l-k)^2\, (l-p)^2\, \bn \!\cdot\! (k-l)}
\end{align}
with $p^2 = (k-p)^2 = 0$ and $\bar{n}$ as defined in Appendix~\ref{sec:TChange}. In light-cone gauge, this integral
appears when using the On-Shell Diagram method calculation for diagrams with the topology of 
\fig{threepoint}a with a gluon line $l-k$. In Feynman gauge, it contributes to diagrams in which the gluon $l-k$ line is connected to the collinear Wilson line on the left side of the cut, as e.g. in \fig{threepoint}b.
Of course, similar integrals are involved in the calculation performed using the Dispersive Method. In that case, the contributions from `real-real' and `real-virtual' cuts cannot be disentangled easily, so it is not possible to isolate a simple term that contains the light-cone divergence for illustration. Nevertheless, the light-cone regulation works the same as described below for the On-Shell Diagram method.

Let us make a change of loop variables to $l' = k - l$. Defining $p' = k-p$ (with 
$p'^2=0$) we have
\begin{align} \label{eq:LCint2}
\int \dfrac{\df^dl'}{(2\pi)^d} \; \dfrac{\bn\!\cdot\! (k-p') }{(k-l')^2\, l'^2\, (l'-p')^2\, \bn \!\cdot\! l'}\,.
\end{align}

Now, this integral is just the integral in eq.~(A.12) from \mycite{Ellis:1996nn} but with $l \to l'$, 
$p \to p'$, $x \to -z/(1-z)$. Effectively we are now taking the leg $p'$ as the `incoming'
on-shell leg, and the leg $p$ as the `outgoing' on-shell leg, but then given the magnitudes
of the momenta, if we want to regard leg $k$ as outgoing as before it has to have a negative 
light-cone momentum fraction.

Using the result in eq.~(A.12) from \mycite{Ellis:1996nn}, we obtain for \eq{LCint2}
\begin{align} \label{eq:basicthreepoint}
&
 \biggl\{ \int_{0}^{-\tfrac{z}{1-z}} \frac{\df y}{y}\, w^{-\epsilon} (1-w)^{-1-\epsilon}\,
{}_2 F_1 \Bigl[ 1+\epsilon,1;1-\epsilon;\frac{w}{(w-1)(1-z)} \Bigr]
\\\nn & \quad
+ 2 \frac{\Gamma^2 (1-\epsilon)}{\Gamma (1-2 \epsilon)}
(1-z)^{-\epsilon} \int_{-\tfrac{z}{1-z}}^{1} \frac{\df y}{y}\, (1-y)^{-1-2\epsilon} \biggl\}
\times \frac{-\img}{16 \pi^2 k^2}
\biggl(\frac{4 \pi}{-k^2} \biggr)^\epsilon
\frac{\Gamma(1+\epsilon)}{\epsilon}\, \frac{-1}{1-z}
\end{align}
where $y = \bar{n} \cdot l' / \bar{n} \cdot p'$ and $w = -y(1-z)/z$ here. The first integral
in curly brackets is regulated by the $w^{-\epsilon}$ factor. The second integral requires further
regulation, because we integrate $1/y$ over the origin, but this is a very simple integral.
We can use any one of the standard regulators in the $1/y$ factor: principal value, multiplying
the integrand by an infinitesimal negative power of $\abs{y}$, adding a small imaginary part of either sign to the
denominator\footnote{In Feynman gauge this corresponds to consistently assigning a $\pm \img \epsilon$ prescription 
to the collinear Wilson line propagators, which in SCET is a priori not fixed by causality. Prescription-dependent 
terms cancel once the complex conjugated (i.e. the left-right mirror) graph of \fig{threepoint} is added.} (as is done e.g.~in \mycite{Gehrmann:2014yya}, section 3.1), cutting out a small symmetric region from the integration either side of the
origin, etc.. After finally setting the regulator to zero, we will get the same result for any regulator.

\section{Triple gluon field strength vertex}
\label{sec:W3vertex}

The Feynman gauge expressions for the vertices with one and two external legs associated with the gluon field strength operator $\cB_{n\perp}^\mu$, which is part of the operator definition of $B_g$ in \eq{Bg_def}, can e.g. be found in \mycite{Berger:2010xi}.
For completeness we also give the Feynman rule for the $\cB_{n\perp}^\mu$ vertex with three external gluons needed in the Feynman gauge calculation of $\cI^{(2)}_{gg}$:
\begin{align}
&\raisebox{-5 ex}{\includegraphics[width=0.3\textwidth]{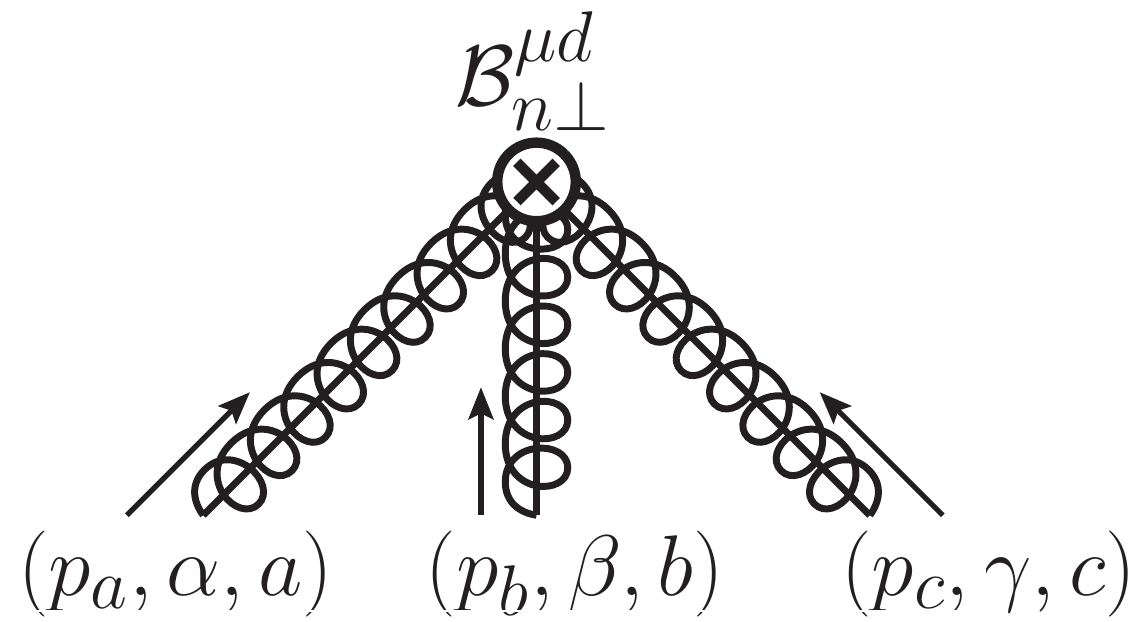}} \hspace{-2 ex}
= g^2\biggl[- g_\perp^{\alpha \mu} \, \bn^\beta \bn^\gamma \bigg( \frac{f^{a\,c\,m}f^{b\,d\,m}}{ \bn \!\cdot\! p_b\; \bn \!\cdot\!(p_b + p_c)}
+ \frac{f^{a\,b\,m}f^{c\,d\,m}}{ \bn \!\cdot\!p_c \; \bn \!\cdot\!(p_b + p_c)} \bigg)  + \text{\rm cycl.} \biggr]
\nn\\[2 ex]
&-g^2\biggl[2 \frac{\bn^\alpha \, \bn^\beta \, \bn^\gamma}{\bn \!\cdot\!p_a \; \bn \!\cdot\!p_c \; \bn \!\cdot\!(p_a+p_b) \;\bn \!\cdot\!(p_b+p_c) \;\bn \!\cdot\!(p_a+p_b+p_c) }
\;\tr\bigl[T^a\, T^b\, T^c\, T^d \bigr]
\label{eq:W3vertex}
\\\nn & \quad\times
\Bigl( p_{a\perp}^\mu\, \bn \!\cdot\!p_a \; \bn\!\cdot\!(p_a+p_b) -  p_{b\perp}^\mu\, \bn \!\cdot\!(p_a+p_b) \; \bn\!\cdot\!(p_b+p_c) + p_{c\perp}^\mu\, \bn \!\cdot\!p_c \; \bn\!\cdot\!(p_b+p_c) \Bigr)
+ \text{perms.} 
\biggr]
\end{align}
Here `cycl.' and `perms.' stand for additional terms generated by cyclic (two more terms) and full permutations (five more terms) in $\{(p_a,\alpha,a),\,(p_b,\beta,b),\,(p_c,\gamma,c)\}$, respectively.
Note that in our two-loop calculation of the partonic beam function with on-shell transverse polarized incoming gluons the second term $\propto \bn^\alpha \, \bn^\beta \, \bn^\gamma$ does not contribute.

\bibliographystyle{../jhep}
\bibliography{../beamfunc}

\end{document}